\documentclass[journal]{IEEEtranTIE}

\usepackage{graphicx}
\usepackage{cite}
\usepackage{picinpar}
\usepackage{amsmath}
\usepackage{url}
\usepackage{flushend}
\usepackage[latin1]{inputenc}
\usepackage{colortbl}
\usepackage{soul}
\usepackage{multirow}
\usepackage{pifont}
\usepackage{color}
\usepackage{alltt}
\usepackage[hidelinks]{hyperref}
\usepackage{enumerate}
\usepackage{siunitx}
\usepackage{breakurl}
\usepackage{epstopdf}
\usepackage{pbox}

\begin{document}
\title{Effects of current limit for grid forming converters on transient stability: analysis and solution}

\author{
	\vskip 1em
	Kanakesh Vatta Kkuni, \emph{Student Member},
 	Guangya Yang, \emph{Senior Member}
	\thanks{The work is supported by Phoenix project, funded by Ofgem under Network Innovation Competition programme, Project Direction ref: SPT / Phoenix / 16 December 2016 (https://www.spenergynetworks.co.uk/pages/phoenix.aspx).
	
	Kanakesh Vatta Kkuni and Guangya Yang are with  are with the Center for Electric Power and Energy, Department of Electrical Engineering, Technical University of Denmark,  2800 Kgs. Lyngby, Denmark  (e-mail: kvkhu@elektro.dtu.dk; gyy@elektro.dtu.dk)
 
	}
}

\maketitle
	
\begin{abstract}
Grid forming control applied to power converters which interface storage or renewable generation to the power grid, has been identified as a potential solution to facilitate a substantial share of converter-based renewable generation in the power system. Analyzing the response of grid forming converters (GFC) for large frequency, phase, and voltage events, particularly when the GFC enters the current limit operation, is very important for system stability, but studies on this have been limited. This paper presents a quantitative and illustrative analysis of the impact of the current limit in GFC on the transient stability of a system comprising of GFC. Furthermore, a solution based on virtual active power is proposed to improve the transient stability margin of the GFC when the GFC enters the current limit. Finally, the analysis and the proposed method to enhance the transient stability are verified by Power hardware in the loop (PHIL) experimental tests.
\end{abstract}


{}

\definecolor{limegreen}{rgb}{0.2, 0.8, 0.2}
\definecolor{forestgreen}{rgb}{0.13, 0.55, 0.13}
\definecolor{greenhtml}{rgb}{0.0, 0.5, 0.0}

\section{Introduction}

\IEEEPARstart{T}he core of the bulk power system is evolving from conventional synchronous machine-based generation (SG) to asynchronous power converter-based renewable generation and energy storage. Ensuring continued reliability of the power system when the fundamental nature of the power systems is altered with a reducing share of synchronous generators is a challenge. The grid forming control for converters (GFC) or virtual synchronous machine control is a potential solution to mitigate the challenges caused due to reduction of SG \cite{ENTSO-E2019,NGSOF}.  GFC's characteristics which are beneficial in ensuring system security are: Higher stability margin, Instantaneous fault current contribution, Inertial response, and instantaneous active power contribution to phase jumps and system frequency changes. 
In recent years several variations of grid forming control have been proposed. The difference between the GFC's is mainly attributed to how the synchronization loop or active power control and voltage control is realized. The implementation has been ranging from simple active power-frequency ($P-f$) and reactive power voltage ($Q-V$) droop control \cite{Paquette2015} based electromechanical realization to a full realization of synchronous machine dynamic equations \cite{zhong2010synchronverters,zhang2016synchronous}. The studies about electromagnetic model realization of GFC have been discussed in \cite{natarajan2017synchronverters,rodriguez2019virtual,mo2016evaluation}.

 The SG has an overload capacity of several times its rated value during short-term overload and grid faults. Such high overload capability is not typical for a power electronic converter-based GFC, which emulates SG. Therefore, GFC needs to have a robust overload and fault current limiter for protecting and ensuring reliable GFC operation during significant transient events. Yet, implementing the power or current limiters for GFC is challenging because an instantaneous response is expected from GFC similar to a voltage source. In addition,  for GFC, the active power, which is influence by current limiting, also acts feedback variable for synchronization. Consequently, there is a possibility of wind up of power-based synchronization loop leading to instability \cite{Paquette2015}. Thus the  current limiter can negatively impact GFC responses, particularly during transient events\cite{Wang2020a,rosso2021grid,Paquette2015}.

Switching to control that employs voltage-based phase-locked loop (PLL) synchronization and controls the active and reactive current control during overload and grid fault scenarios is an option for GFC \cite{shi2017low}. However, the solution is not robust as there is a need to preserve a complete additional control set in the same digital control. Furthermore, the stability issues caused due to PLL is also undesirable \cite{Wang2020a}. A backup PLL is not required in the current limit methods employed in \cite{Taul2020,huang2017transient} or with the virtual impedance-based current limit method discussed in \cite{Paquette2015,Qoria2020b}.  However, it is important to evaluate the transient stability of the GFC equipped with such current limit methods for all possible transient events. 

 Recently many studies have analyzed the transient stability of GFC \cite{Wu2019,Wang2020a,ma2020virtual,huang2017transient}. Some of these studies were conducted on GFC with no current limit implemented \cite{Wu2019,Wang2020a}. A current limit is likely to be triggered in a practical transient event scenario. Therefore studying the GFC for transient events requires the inclusion of the current limit for a practical understanding. The transient stability of a droop-based grid forming converter with the current limit is studied in \cite{ma2020virtual,huang2017transient} and methods are discussed for improving the transient stability, however, discussed approaches are not intuitive and requires tuning of parameters. In addition, all the transients studies have been restricted only to a voltage dip event when connected to an infinite voltage source. A GFC, when operated in a microgrid in parallel with an SG and virtual impedance current limit, can avoid the wind-up effect of the outer loop for load change events \cite{Paquette2015}.  However, only a limited evaluation of droop control-based GFC only for load change events has been shown. The frequency and phase jump events are significant in the present context for the power system with a significant renewable share and low inertia. Therefore transient events such as a change of frequency and phase jump can cause current limit triggering for GFC. A detailed study on GFC under such transient events or in the presence of an SG is missing in the literature. Furthermore, a quantitative analysis of the transient stability of GFC for all the possible events is also missing literature.

This paper first demonstrates the scenarios in which the current limit could be triggered in a GFC working in a low inertia power system and outlines the challenges in maintaining the GFC synchronism with the rest of the grid when the current limit is triggered for each scenario. Then, to clarify the stability phenomena when the current limiter is activated, a quantitative analysis of transient stability for events such as phase jump and frequency change events and voltage dip events when the current limiting algorithm is employed in a GFC is presented in this paper. The transient stability margins such as maximum phase jump, maximum rate of change of frequency, critical fault clearing time, and methods to improve the synchronization stability for all the three events described in this paper are detailed from the conducted analysis. Finally, a coordinated overload and current limit for grid forming converter employing virtual power, extending the transient stability margin for all the grid events, are proposed in this paper. Conducted analysis and the proposed solution to improve the transient stability of the GFC is validated using power hardware in the loop (PHIL) experimental tests are presented. The PHIL evaluation is conducted in a simplified equivalent circuit and a modified lower inertia IEEE 9 bus system.
 
\section{Grid Forming Converter Configuration}
\begin{figure}[h]
\centering
    \includegraphics[width=8.5cm]{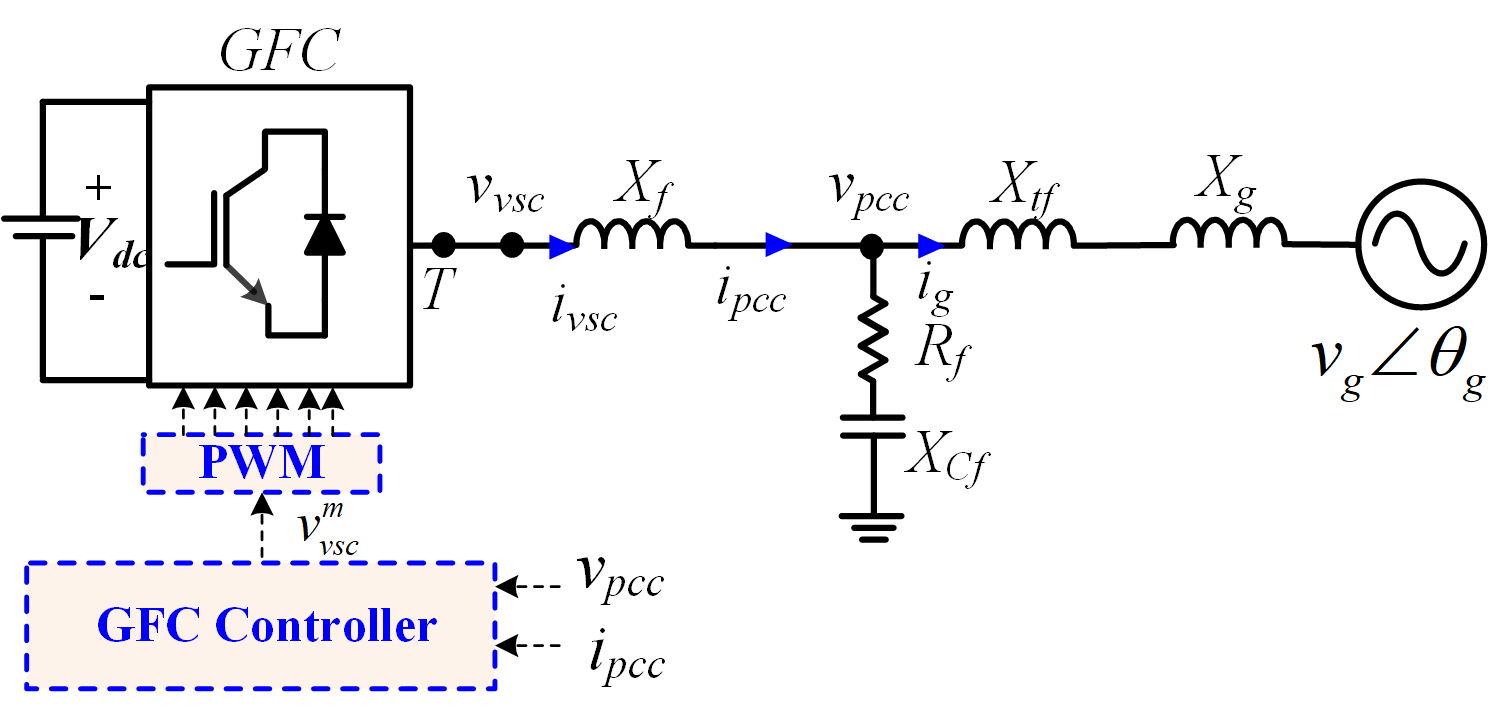}
    \caption{The configuration of 3 phase grid connected GFC system}
    \label{fig:GFC_system}
\end{figure}

\begin{figure}[h]
    \centering
    \includegraphics[width=8.5cm]{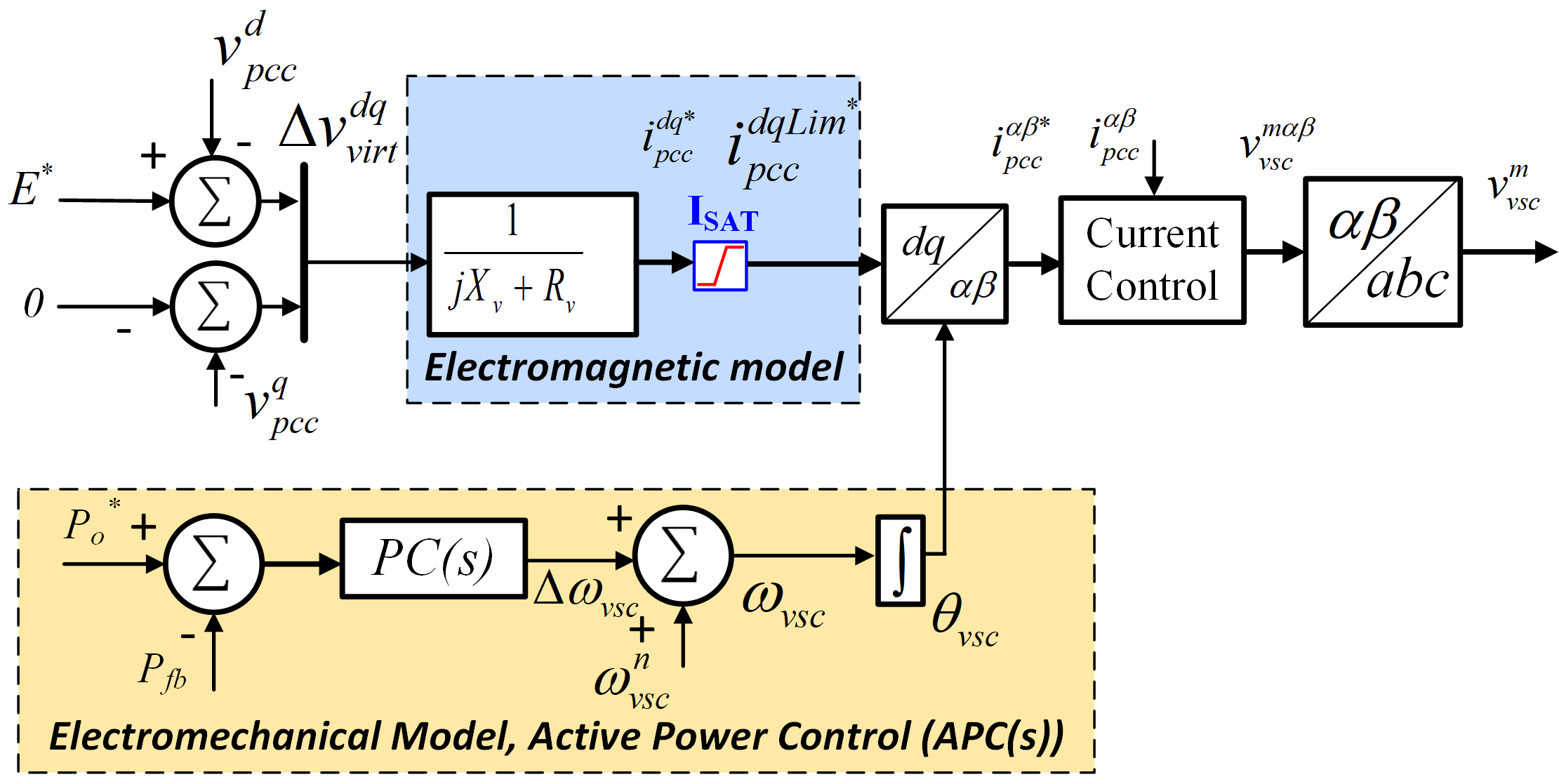}
    \caption{General Control System of GFC with current control inner loop and virtual admittance}
    \label{fig:spc_v1}
\end{figure}

The single line diagram of the GFC considered in the paper is shown in Fig. \ref{fig:GFC_system}. The filter circuit comprises reactor $X_f$, capacitor of reactance $X_{Cf}$, and damping resistor $R_f$. The point of common coupling (PCC) is at the terminal of the filter capacitor, where measurements $v_{pcc}$, $i_{pcc}$, and $i_{g}$ are taken. The reactance $X_{tf}, X_g$ represents the reactance of transformer and grid impedance, respectively. The grid forming control with an inner current control loop with current references derived from virtual admittance is used in this paper is as shown in Fig. \ref{fig:spc_v1} \cite{zhang2016synchronous,zhang2017frequency,mo2016evaluation,Taul2020}. The power control system consists of inertia, frequency droop, and damping emulation through active power control ($APC(s)$) with active power setpoint ($P_o^*$). The power output from the GFC measured at the PCC ($P_{pcc}$) is typically chosen as the feedback power ($P_{fb}$) to the active power control loop. 
A lead-lag (LL) compensator-based power controller ($PC(s)$) is employed to emulate electromechanical behaviour of the synchronous machine \cite{zhang2017frequency}. The compensator is given as
\begin{equation}
\Delta \omega_{vsc}=\underbrace{\frac{K_{pp}s+K_{ip}}{s+K_{gp}}}_\text{PC(s)}*(P_o^*-P_{fb})
\label{eqn:LL_PC}
\end{equation}
 The compensator $PC(s)$, similar to the swing equation-based electromechanical model can realize virtual inertia constant, H, and frequency droop $R_d$. In addition, LL-based electromechanical model has an additional degree of freedom to increase the power control's damping coefficient ($\zeta$). The parameters ($K_{pp},K_{ip},K_{gp}$) can be calculated from a given virtual inertia constant H in seconds, power-frequency droop gains $R_d$ in pu and damping coefficient ($\zeta$). 
\begin{equation}
   K_{ip}=\frac{\omega_B}{2*H},K_{gp}=\frac{K_{droop}}{2*H} \text{ where } K_{droop}=\frac{1}{R_d}
    \label{eqn:LL_PC2}
\end{equation}

\begin{equation}
 K_{pp}=\zeta \sqrt{\frac{2 \omega_B}{P_{max}H}}-\frac{K_{droop}}{2*H*P_{max}}
 \label{eqn:LL_PC3}
\end{equation}
Where $\omega_B$ is the base frequency of the system, $P_{max}$ is the maximum static power transfer possible between the GFC and an infinite voltage source. If the power-frequency droop is not required, the parameter $K_{droop}$ is to be set to zero in (\ref{eqn:LL_PC2})-(\ref{eqn:LL_PC3}).
For the GFC electromagnetic model, a quasi-stationary model has been shown superior to a dynamic electromagnetic model model for GFC \cite{mo2016evaluation}. Variable superscripted with d-q are variable vectors of the direct and quadrature frame original parameters represented in the synchronously rotating reference frame defined by the virtual rotor of the GFC ($\theta_{vsc}$).
The electromagnetic model consists of the internal voltage source ($E$) in series with an algebraic representation of an impedance. It is realized by multiplying the difference between the internal voltage source ($E^*$) of the GFC and PCC voltage ($v_{pcc}^{dq}$) with virtual phasor admittance resulting in reference unsaturated stator current ($I^{dq*}_{pcc}$)
\begin{equation}
   I^{dq*}_{pcc}=\frac{E-v_{pcc}^{dq}}{R_v+jX_v}
    \label{eqn:zin}
\end{equation}
where $R_v, X_v$ are the virtual internal virtual resistance and reactance of the GFC. The virtual impedance is chosen such that the output impedance is predominantly inductive with an X/R ratio of 10. The electromagnetic model of the GFC also includes the current limiting algorithm. Fast acting current limiters for GFC is critical as the GFC response to grid events is nearly instantaneous. A circular current limit on $I^{dq*}_{pcc}$, as shown in (\ref{eqn:Ilim}) is an ideal choice as it precisely limits current and preserves the angle of the injected current and thus limiting the interaction with the active power-based synchronization \cite{gkountaras2015evaluation,Taul2020}.

The limited current vector $i_{pcc}^{dqLi{m^*}}$ is given by 
 \begin{equation}
i_{pcc}^{dqLi{m^*}}=\frac{1}{KC_{lim}}*I^{dq*}_{pcc},\text{ where }   KC_{lim}= \frac{\left| I^{dq*}_{pcc} \right|}{I^{Lim}}
\label{eqn:Ilim}
\end{equation}

$\left| I^{dq*}_{pcc} \right|$ is the magnitude of the unsaturated reference current vector and is equal to $\sqrt{(I^{d*}_{pcc})^2+(I^{d*}_{pcc})^2}$
and $I^{Lim}$ is the nominal maximum peak current, and thus the vector $i^{dqLim^*}_{pcc}$ is of the magnitude $I^{Lim}$ during current limited operation.

The paper's focus is on the interaction between the current limit and active power control and its consequences on the transient stability of GFC. Therefore, to preserve the paper's brevity, the outer reactive power control or voltage control is assumed to be slow. Thus, the voltage control and reactive power control dynamics are not accounted for in this work. Also, the DC link dynamics and other supervisory controls are not considered in this paper for the sake of easier understanding.

\section{GFC analysis under current limit}
The analysis assumes that the dynamics of the employed current limit are faster than other control and thus time scale separated from the stability study considered in the paper. Additionally, the paper's focus is applying the GFC on large power systems with a low R/X ratio; therefore, only the reactances of the virtual impedances and grid impedances are considered in the quasistatic analysis.

\begin{figure}[h]
    \centering
    \includegraphics[width=8.5cm]{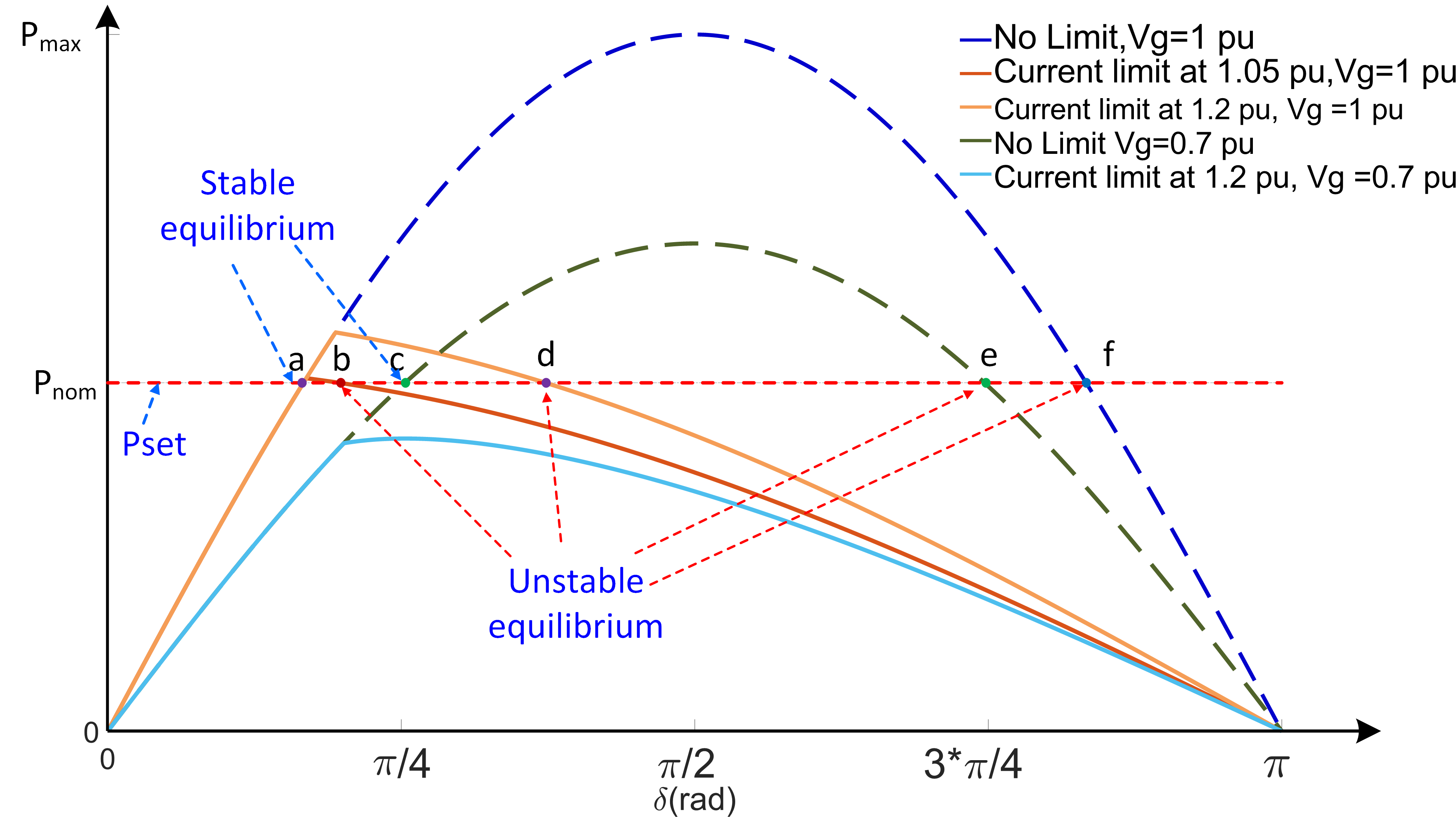}
    \caption{Power angle curves for GFC with and without limiter activation at different grid voltage conditions}
    \label{fig:equal_area_1}
\end{figure}

The active power ($P_{GFC}$) and the reactive power ($Q_{GFC}$) at the internal voltage source ($E$) is given by 
 \begin{equation}
P_{GFC}=\frac{E*Vg*\sin{\delta}}{X_T},
Q_{GFC}=\frac{E^2-E*Vg*\cos{\delta}}{X_T}
\label{eqn:PQ_GFC}
\end{equation}
Where $X_T$ is the total reactance of the system from between the internal voltage ($E$) and is equal to sum of $X_{tf}$, $X_g$ and $X_v$, and the infinite voltage source and $\delta$ is the angle difference between the internal voltage source and infinite voltage source and is equal to difference between $\theta_{vsc}$ and $\theta_{g}$.
The active and reactive power at the internal virtual voltage source $E$ in the rotating dq frame defined by GFC virtual angle ($\theta_{vsc}$) can also be written as
 \begin{equation}
P_{GFC}=E*I^{d*}_{pcc},
Q_{GFC}=-E*I^{q*}_{pcc}
\label{eqn:PQ_GFC_dq}
\end{equation}
The magnitude of the current vector $I^{dq*}_{pcc}$ can then also be written as
\begin{equation}
\left| I^{dq*}_{pcc} \right|=\frac{\sqrt{P_{GFC}^2+Q_{GFC}^2}}{E}=\frac{M_v}{X_{T}}
\label{eqn:I_mag2}
\end{equation}
where $M_v$ is equal to $\sqrt{Vg^2+E^2-2*Vg*E*\cos(\delta)}$
The generalized power transfer equation for the GFC can then be calculated from (\ref{eqn:Ilim}),(\ref{eqn:PQ_GFC}),(\ref{eqn:I_mag2})
\begin{equation}
P_{GFC} = \left\{ {\begin{array}{*{20}{c}}
{\frac{{E*Vg*\sin \delta }}{{{X_T}}},\hspace{20mm}\left| {{i_{pcc}}} \right| \le {I^{Lim}}}\\
{\frac{{E*Vg*\sin \delta }}{{\sqrt {V{g^2} + E{^2} - 2E*Vg*\cos (\delta )} }}*{I^{Lim}},{\rm{ }}\left| {{i_{pcc}}} \right| > {I^{Lim}}}
\end{array}} \right.
\label{eqn:PQ_gen}
\end{equation}
It can be observed that the power transfer under current limited case is independent of the network reactance. When the drop across the virtual resistance ($R_v$) is neglected the active power at the internal voltage terminal is same as $P_{pcc}$. The power angle curves for the GFC with and without limiter activation at different grid voltage conditions are shown in Fig. \ref{fig:equal_area_1}. It could be seen that the unstable operating points occurs at much lower phase angle ($\delta$) and maximum power transfer possible is greatly reduced when under current limit.

\subsection{Internal impedance of GFC during current limit}
Recalling from (\ref{eqn:zin}) and (\ref{eqn:Ilim}) under current limit case is equivalent to the internal impedance $Z_{in}^{cc}$ as shown below
\begin{equation}
    Z_{in}^{cc}=KC_{lim}*(R_v+jX_v)
\label{eqn:Zlim_cc}
\end{equation}

The value for $KC_{lim}$ is automatically set due to control objective in (\ref{eqn:Ilim}) to ensure the value of peak value of the GFC current is equal to $I^{Lim}$. Therefore for a given sustained disturbance at the PCC the, the input reactance ($X_{in}^{^{cc}}$) for the current controlled GFC in current limited case can be written as
\begin{equation}
    X_{in}^{cc}=K_{lim}*X_v
\end{equation}
A virtual impedance based current limiting increases the internal impedance of the GFC for limiting the current \cite{Paquette2015}. For a given sustained disturbance, the internal impedance of GFC with voltage control equipped with virtual impedance based current limiting should be the same as in the case of GFC structure considered in this paper with current control and circular current limiting. This is because both the current limiting methodologies have similar objective of keeping the output current below nominal value without directly changing the phase of injected current. Hence, the analysis and conclusions derived from this paper based on GFC structure with current control and circular current also holds true for GFC employed with virtual impedance current limit.

\section{Transient stability Evaluation of GFC}
The synchronous generator (SG) transient stability  focusing on rotor dynamics is evaluated typically for fault cases. If the rotor angle of the SG is not diverging with the rest of the system after the fault event, the SG is stable against transient events. The transient stability of GFC with no current limit during under-voltage dip events has been extensively studied \cite{Wang2020a,Wu2019}. However, GFC's synchronization ability has not been evaluated for other large system transients such as phase jumps. Furthermore, few studies have investigated the response of GFC with the current limit during large disturbances. 
\begin{figure}[h]
    \centering
    \includegraphics[width=8.5cm]{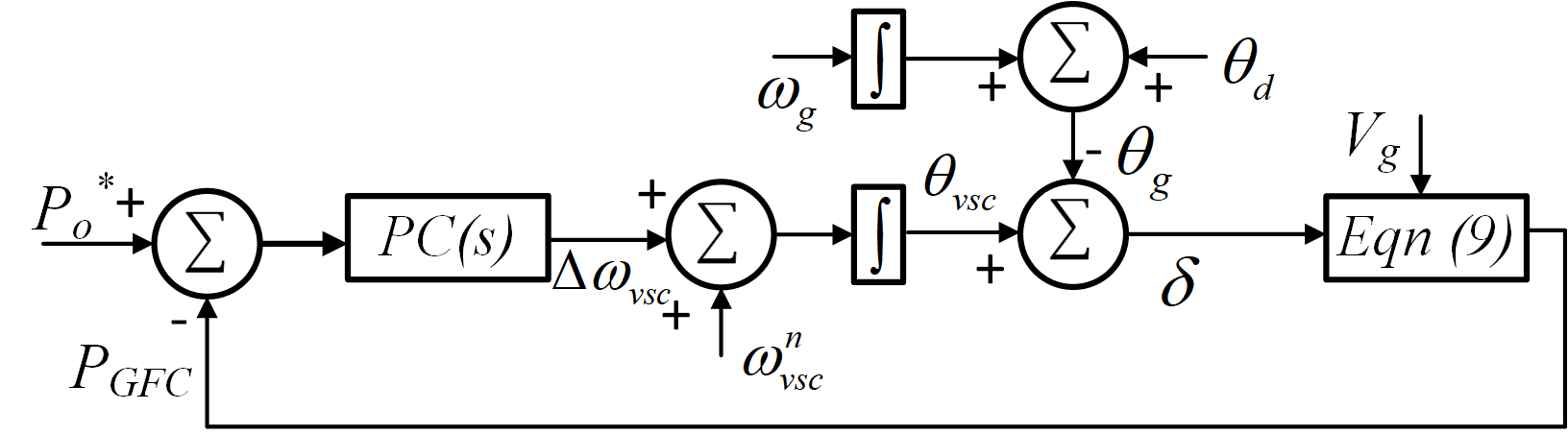}
    \caption{Simplified quasi-static GFC connected to infinite bus}
    \label{fig:GFC_quasi_2}
\end{figure}
In this section, the transient performance of GFC with adn without the current limit implemented is analyzed during system events such as voltage dips, phase shift, and a RoCof event, and conditions for GFC to maintain synchronism for each event is qualitatively described. All events typically co-occur but are studied separately here for an intuitive understanding. The quasi-static model shown in Fig. \ref{fig:GFC_quasi_2} is solved numerically. The total reactance ($X_T$), including the internal reactance of 0.3 pu, is chosen at 0.5 pu, both the internal GFC voltage and infinite bus voltage are assumed to be 1 pu for the analysis, and the current limit is set to be 1.1 pu. The damping coefficient ($\zeta$) has been set to 0.4 to study underdamped control.
\subsection{GFC response against RoCoF}
\begin{figure}[h]
    \centering
    \includegraphics[width=8.5cm]{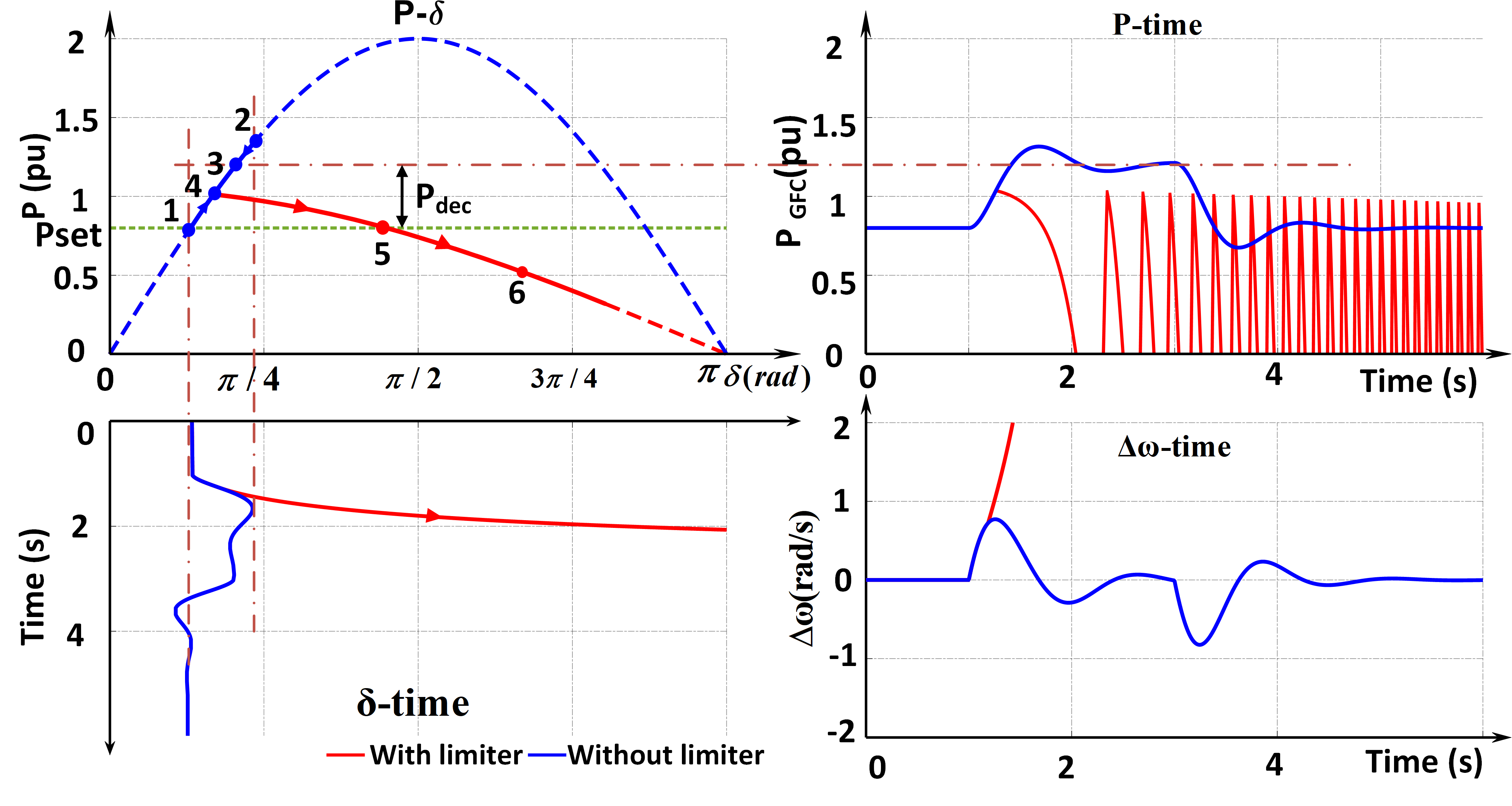}
    \caption{GFC response for a 1 Hz/s ROCOF}
    \label{fig:GFC_rocof}
\end{figure}

The infinite voltage source frequency $\omega_g$ is varied from 50 Hz to 48 Hz at a rate of change of frequency (RoCoF)-1 Hz/s. 
The response of the GFC against the specified RoCof with inertia constant H of 10 s is shown in Fig. \ref{fig:GFC_rocof}. 

When the infinite voltage source is decelerating, a similar deceleration is required from the virtual rotor of the GFC to stay in synchronism. The necessary deceleration power in pu ($P_{dec}$) for the virtual GFC  rotor is given by
\begin{equation}
    P_{dec}=-1*\frac{2*H}{f_{vsc}^n}*RoCof=0.4 pu
    \label{eqn:rocof}
\end{equation}
Where the $f_{vsc}^n$ is the nominal frequency of the GFC. 
When the GFC with no current limits and operating at a power setpoint of 0.8 pu with an operating point at point 1 (Fig. (\ref{fig:GFC_rocof})) is presented with a RocoF event, the rotor angle moves in the trajectory 1-2-3 and settles at point 3 such that a deceleration power of 0.4 pu is impressed on the virtual rotor according to (\ref{eqn:rocof}). However, point 3 is not reachable for the current limited case as shown in Fig. \ref{fig:GFC_rocof}. Therefore the trajectory on the power angle curve  with current limit follows 1-4-5, and beyond point 5, an unstable operating point is reached, and synchronism is lost, which is reflected in all the curves in Fig. \ref{fig:GFC_rocof}. 
When the GFC is also programmed to provide frequency droop ($Rd$) the condition in (\ref{eqn:rocof}) is modified to   
\begin{equation}
    P_{dec}=-1*\frac{2*H}{f_{vsc}^n}*RoCof+\frac{\Delta \omega_{vsc}}{Rd*\omega_{vsc}^n}
    \label{eqn:rocof2}
\end{equation}

From (\ref{eqn:rocof2}) one can also conclude that dynamically reducing $P_{set},H,Rd$ are some of the options to ensure $P_{dec}$ on the virtual rotor and maintain synchronism.
\subsection{GFC response against phase jump}
\begin{figure}[h]
    \centering
    \includegraphics[width=8.5cm]{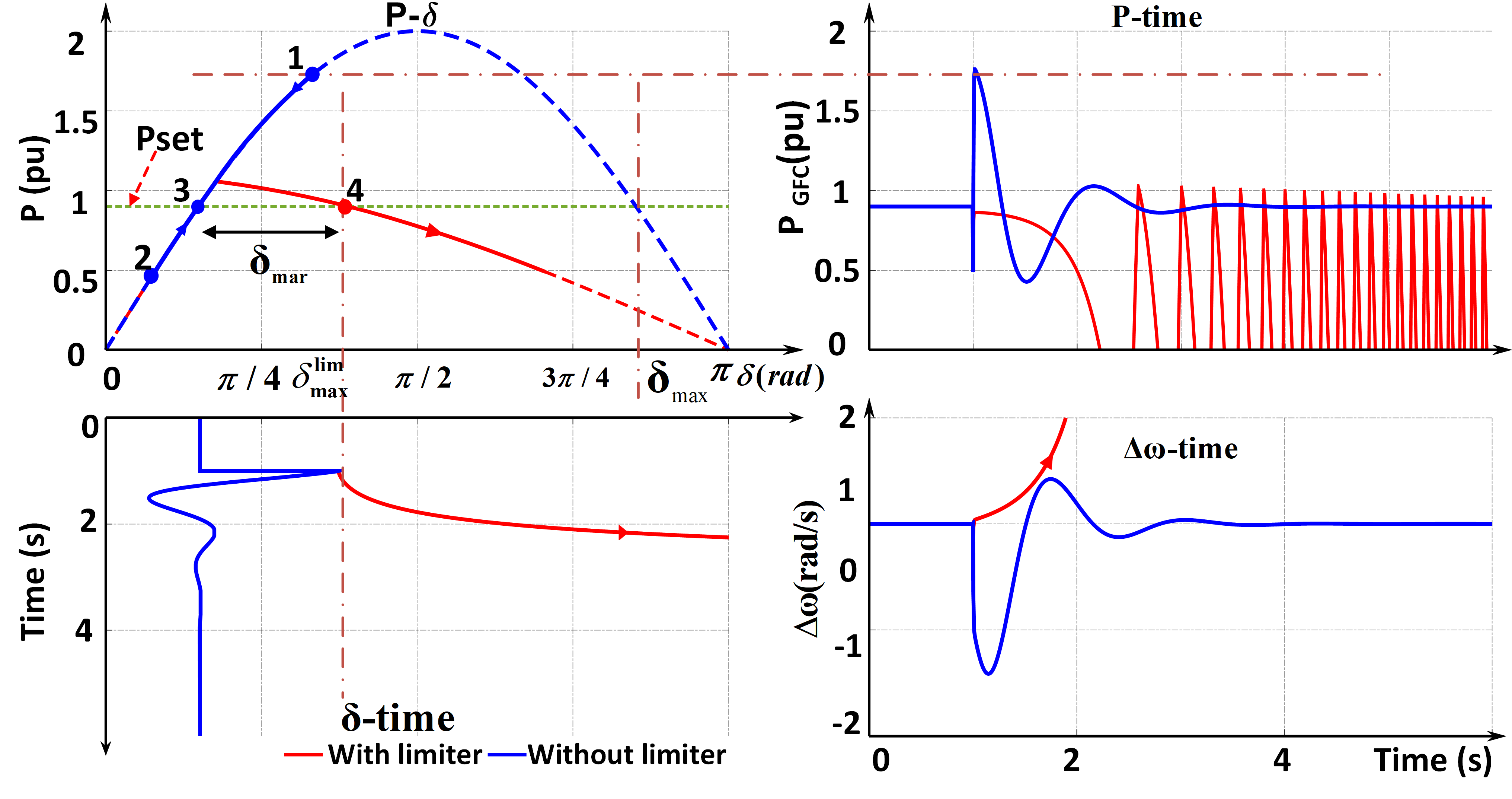}
    \caption{GFC response for a phase jump of 40 deg}
    \label{fig:GFC_phase_shift_40}
\end{figure}
The GFC is expected to respond instantly to phase jumps in line with a voltage source behavior expected of GFC. The GFC response to simulated infinite voltage phase jump of $40^\circ$ is shown in Fig. \ref{fig:GFC_phase_shift_40} with power set point at 0.9 pu $P_{set}$. The grid phase jump is simulated by changing $\theta_d$ in Fig. \ref{fig:GFC_quasi_2}. From the power angle curve ($P-\delta$) of current limited and no limit cases, it is seen that the maximum phase shift between the infinite voltage source and GFC internal voltage is significantly reduced from $\delta_{max}$ to $\delta_{max}^{lim}$. For the simulated phase shift, the trajectory without limiter shifts from point 3 to 1 instantly and then falls back to 3. Whereas for GFC with the current limit, the trajectory instantly moves from 3-4 and thus marginally beyond the stable operating point 4 and loses synchronism. For a given $P_{set}$ the maximum phase shift margin possible ($\delta_{mar}$) can be solved from (\ref{eqn:PQ_gen}).  

The inertia and damping parameters do not impact the stability margin as the instability expected is the instantaneous response.

\subsection{GFC response voltage dip}
The event simulates a power system fault case. The infinite bus voltage is reduced to 0.5 pu for 0.3 s with a $P_{set}$ at 0.8 pu. The power angle curve reduces in magnitude with the reduction in infinite bus voltage, causing the virtual rotor to accelerate. For the unlimited case, the power angle trajectory is 1-2-5-7 and back to 1. The conventional equal area criterion for transient stability is applicable in this case, and the acceleration area 1-2-5-7 is well less than the decelerating area 5-7-9. However, the accelerating area 1-2-3-5 is higher for the current limited case than the available decelerating area 5-6-8 and the momentum gained during acceleration results in rotor angle crossing the unstable equilibrium point 8 and instability.    

\begin{figure}[h]
    \centering
    \includegraphics[width=8.5cm]{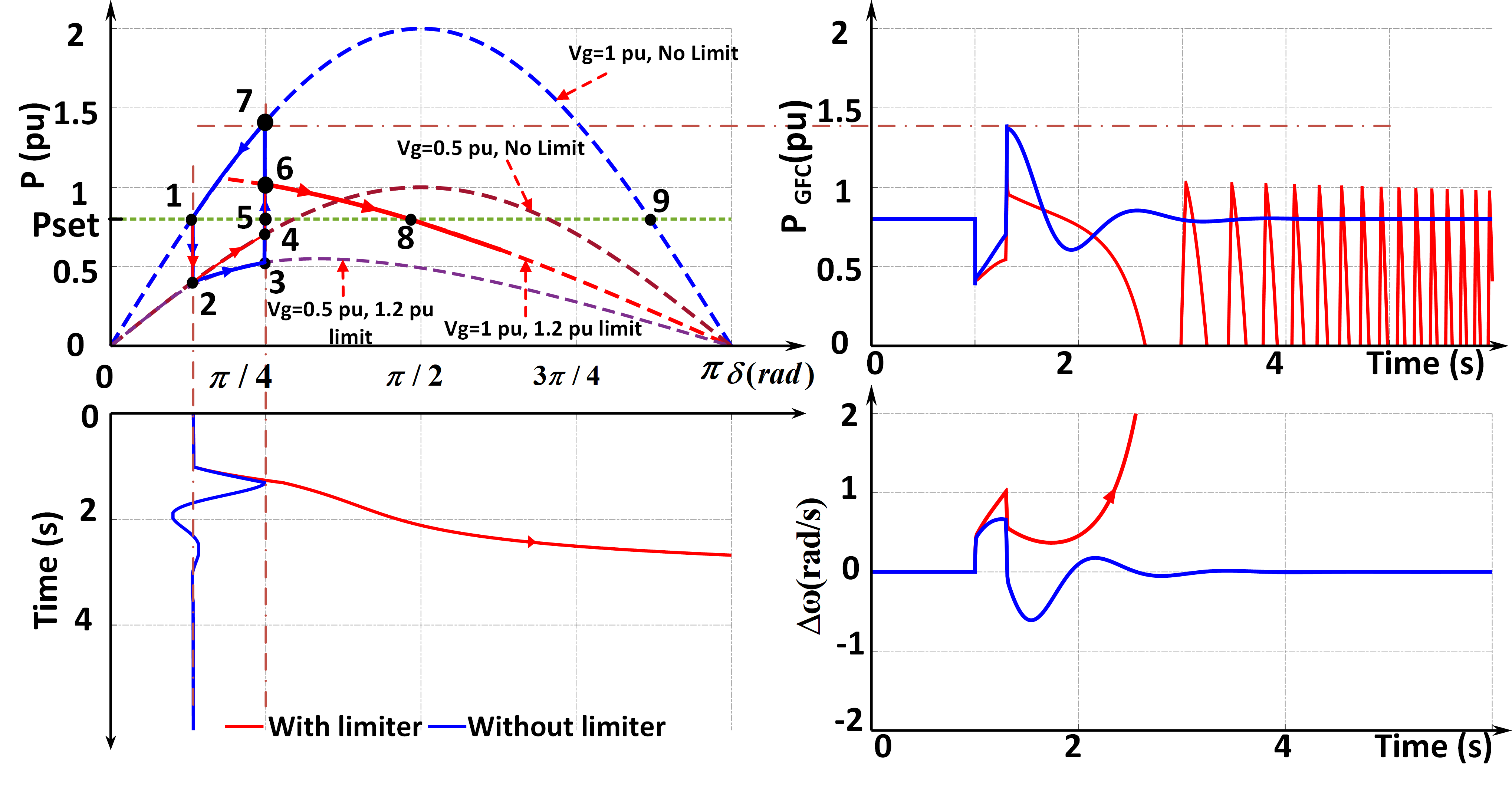}
    \caption{GFC response for a voltage dip of 0.5 pu }
    \label{fig:GFC_fault}
\end{figure}

Most of the existing studies have only evaluated the transient stability of the GFC for fault cases. In practise, GFC is not a rotational rigid body as SG, wherein it is difficult to achieve the mechanical power input reduction during fault cases through fast valving or by dynamic braking resistors, thereby limiting the acceleration torque. Both fast valving and dynamic braking can be easily implemented in a GFC by dynamically changing the power reference or simply freezing the power-based synchronization for a short duration \cite{pattabiraman2020transient,roscoe2016vsm}, provided there are sufficient capacity for energy dissipation. Another possibility is to increase  P-F droop $Rd$, increase programmed inertia H, the stability margin improves as the rotor acceleration is reduced with higher inertia and damping, unlike the RoCof case.

\section{Proposed virtual power based GFC}
The challenge with saturating current is that the active power output becomes insensitive to change of phase, thereby rendering the synchronization control ineffective. In addition methods such as dynamically changing $H$ and $R_d$ is complicated and contradictory for different transient events as explained in the previous section. To counter such limitation, this paper proposes utilizing the unsaturated current references ($I_{pcc}^{dq*}$) for power measurements for the synchronization loop instead of active power measurements at the PCC, i.e. the active power feedback to $APC(s)$ is chosen as 
\begin{equation}
    P_{fb}=P_{virt}=\frac{3}{2}(V_{pcc}^d.I_{pcc}^{d*}+V_{pcc}^q.I_{pcc}^{q*})
\end{equation}
One could also measure the virtual unsaturated power at the GFC virtual internal voltage terminals. Utilizing the virtual unsaturated power as the feedback or controlled parameters for the power synchronization is analysed in this section.

\begin{figure}[h]
    \centering
    \includegraphics[width=8.5cm]{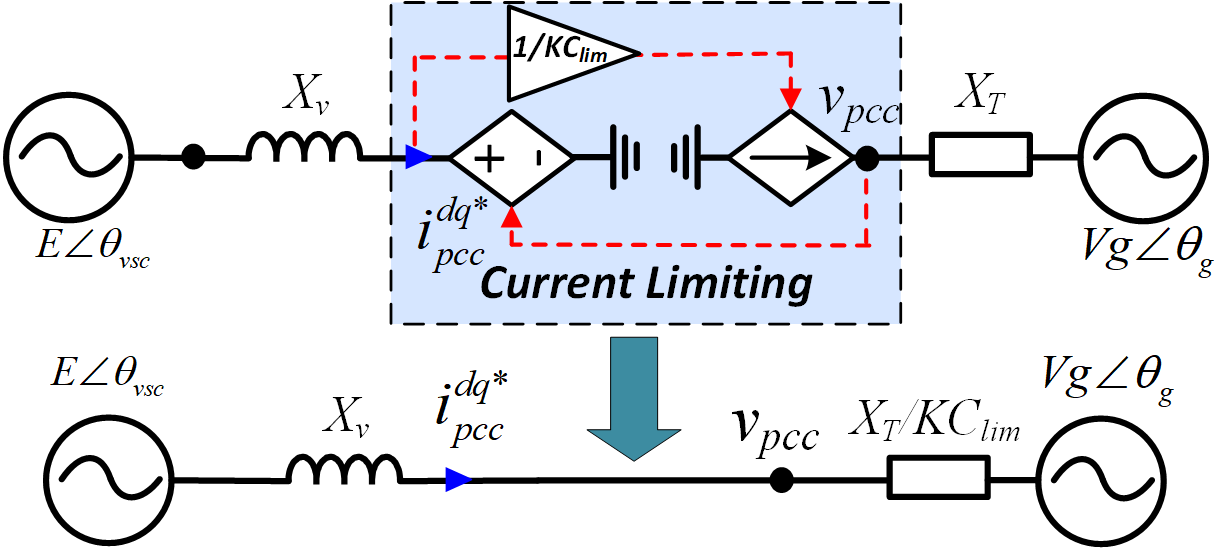}
    \caption{Electrical equivalent circuit of the GFC with unsaturated active power feedback as synchronisation during current limit}
    \label{fig:eqv_current_limit}
\end{figure}
\subsection{Analysis of virtual power based GFC}
For calculation of the virtual power, the unsaturated current references are used, whereas the GFC actual output current when GFC enters the current limit is the saturated current defined by Eq. (\ref{eqn:Ilim}). When the GFC is not in the current limit operation, there is no difference between the $P_{virt}$ and $P_{pcc}$ as long as the current control dynamics are neglected. In practice, the current controller is much faster than the power control loop, and thus the impact of current control dynamics on the choice of power feedback is minimal. Hence an equivalent electrical circuit as shown in Fig. \ref{fig:eqv_current_limit} can be drawn to represent the GFC employed with virtual power feedback. During the current limit operation with virtual power feedback, dividing the current reference by $KC_{lim}$ is equivalent to dividing the net grid side reactance by $KC_{lim}$ as shown in Fig. \ref{fig:eqv_current_limit}.

\begin{figure}[h]
    \centering
    \includegraphics[width=8.5cm]{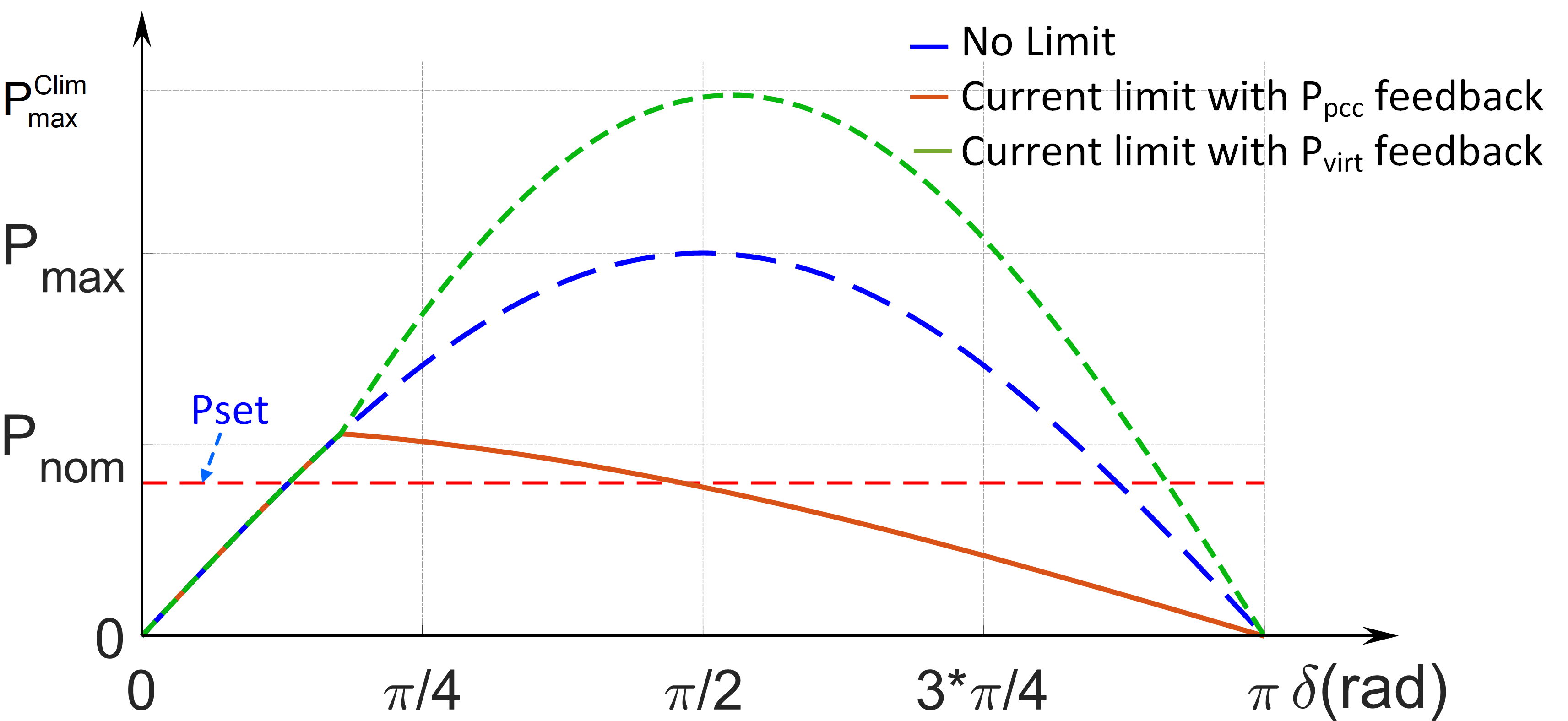}
    \caption{Power angle curves for GFC with and without limiter activation with virtual unsaturated power feedback for power synchronism loop}
    \label{fig:equal_area_2}
\end{figure}

\begin{figure}[h]
    \centering
    \includegraphics[width=8cm]{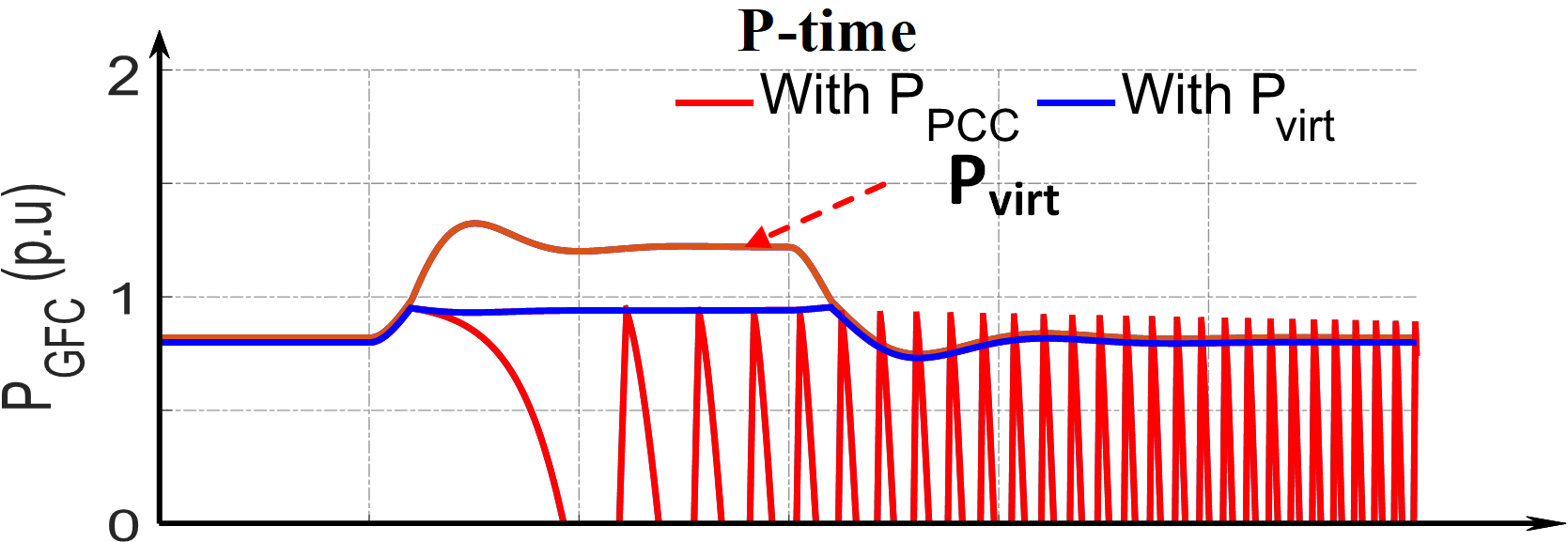}
    \caption{GFC response for a 1 Hz/s ROCOF with virtual power feedback}
    \label{fig:GFC_rocof_pvirt}
\end{figure}

\begin{figure}[h]
    \centering
    \includegraphics[width=8cm]{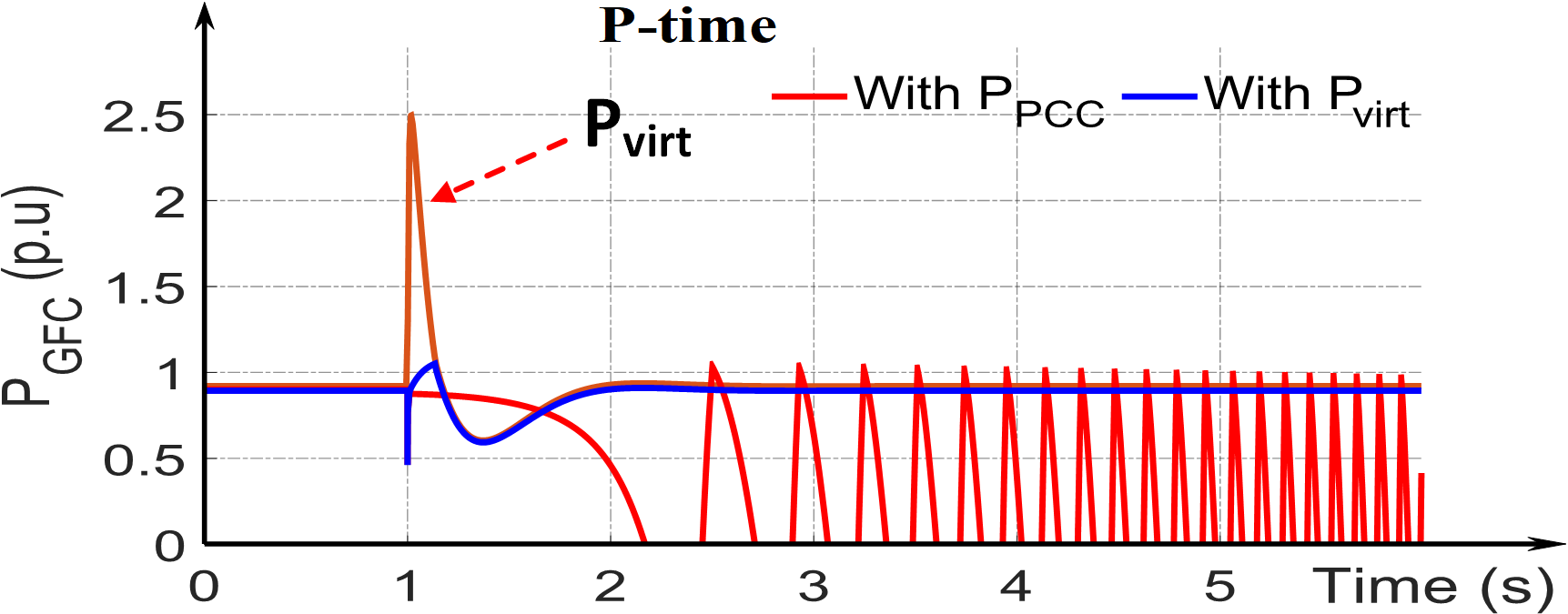}
    \caption{GFC response for a phase jump of 40 deg with virtual power feedback}
    \label{fig:GFC_phase_shift_40_pvirt}
\end{figure}

\begin{figure}[h]
    \centering
    \includegraphics[width=8cm]{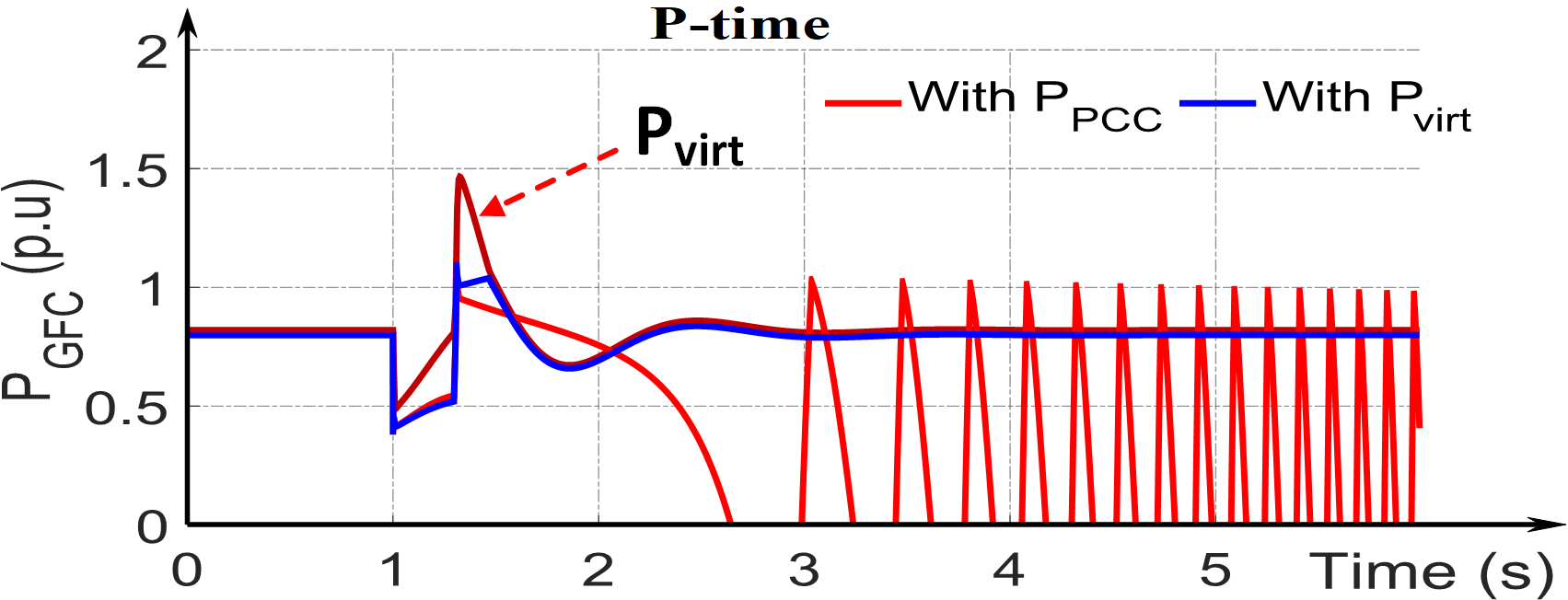}
    \caption{GFC response for a voltage dip of 0.5 pu with virtual power feedback}
    \label{fig:GFC_fault_pvirt}
\end{figure}

From the equivalent circuit shown in Fig. \ref{fig:eqv_current_limit} the unsaturated active and reactive power when when $\left| {{i_{pcc}}} \right| > {I^{Lim}}$ can be calculated
 \begin{equation}
P_{unsat}=\frac{E*Vg*\sin{\delta}}{Xv+ X_T/ KC_{lim}},
Q_{unsat}=\frac{E^2-E*Vg*\cos{\delta}}{Xv+ X_T/ KC_{lim}}
\label{eqn:PQ_GFC_pvirt1}
\end{equation}

The $KC_{lim}$ can be calculated similar to (\ref{eqn:I_mag2})

\begin{equation}
    KC_{lim}= \frac{\sqrt{Vg^2+E^2-2*Vg*E*\cos(\delta)}}{Xv+ X_T/ KC_{lim}}*1/I^{lim}
\end{equation}
simplifying one can write 
\begin{equation}
    KC_{lim}= (\frac{Mv}{I^{lim}}-X_T)/Xv
\end{equation}

Thus virtual unsaturated power when $\left| {{i_{pcc}}} \right| > {I^{Lim}}$ for can be simplified and written as independent of $KC_{lim}$ as  
 \begin{equation}
P_{virt}=P_{unsat}=\frac{E*Vg*\sin{\delta}}{X_v+Xv.XT/(Mv/I^{lim}-XT)}
\label{eqn:PQ_GFC_pvirt2}
\end{equation}

From (\ref{eqn:PQ_GFC_pvirt1}) and (\ref{eqn:PQ_GFC_pvirt2}) one can easily observe that the current limit activation inherently extend the peak of the power angle characteristics when virtual power $P_{virt}$ is utilized as the feedback variable even beyond the power transfer limit for unlimited case. This can be verified by plotting the power angle curve for GFC with virtual power feedback as shown in Fig. \ref{fig:equal_area_2}. This feature greatly increases the synchronization stability margin of all the cases discussed in the previous section.

The cases studies conducted in the previous section are repeated with virtual power feedback and the results are as depicted Fig. \ref{fig:GFC_rocof_pvirt}-\ref{fig:GFC_fault_pvirt}. As shown in the figures the GFC with virtual power feedback when in current limited operation can sustain all the large transient events discussed in this paper.

   \begin{table}[!t]
	\renewcommand{\arraystretch}{1.3}
	\caption{PHIL Scaling for the VSC hardware}
	\centering
	\label{table_3_2}
	\resizebox{\columnwidth}{!}{
		\begin{tabular}{l l l l}
			\hline\hline \\[-3mm]
			\multicolumn{1}{c}{Symbol} & \multicolumn{1}{c}{Description} & \multicolumn{1}{c}{Physical Value}  & \multicolumn{1}{c}{Scaled to Simulation } \\[1.6ex] \hline
			$V_{vsc}$ & Amplifier voltage &123 V & 12.3kV\\
			$P_{vsc}$ & VSC power & 1.5 kVA & 70 MVA\\
			\hline\hline
		\end{tabular}
	}
\end{table}

\begin{figure}[!t]
    \centering
    \includegraphics[width=8.5cm]{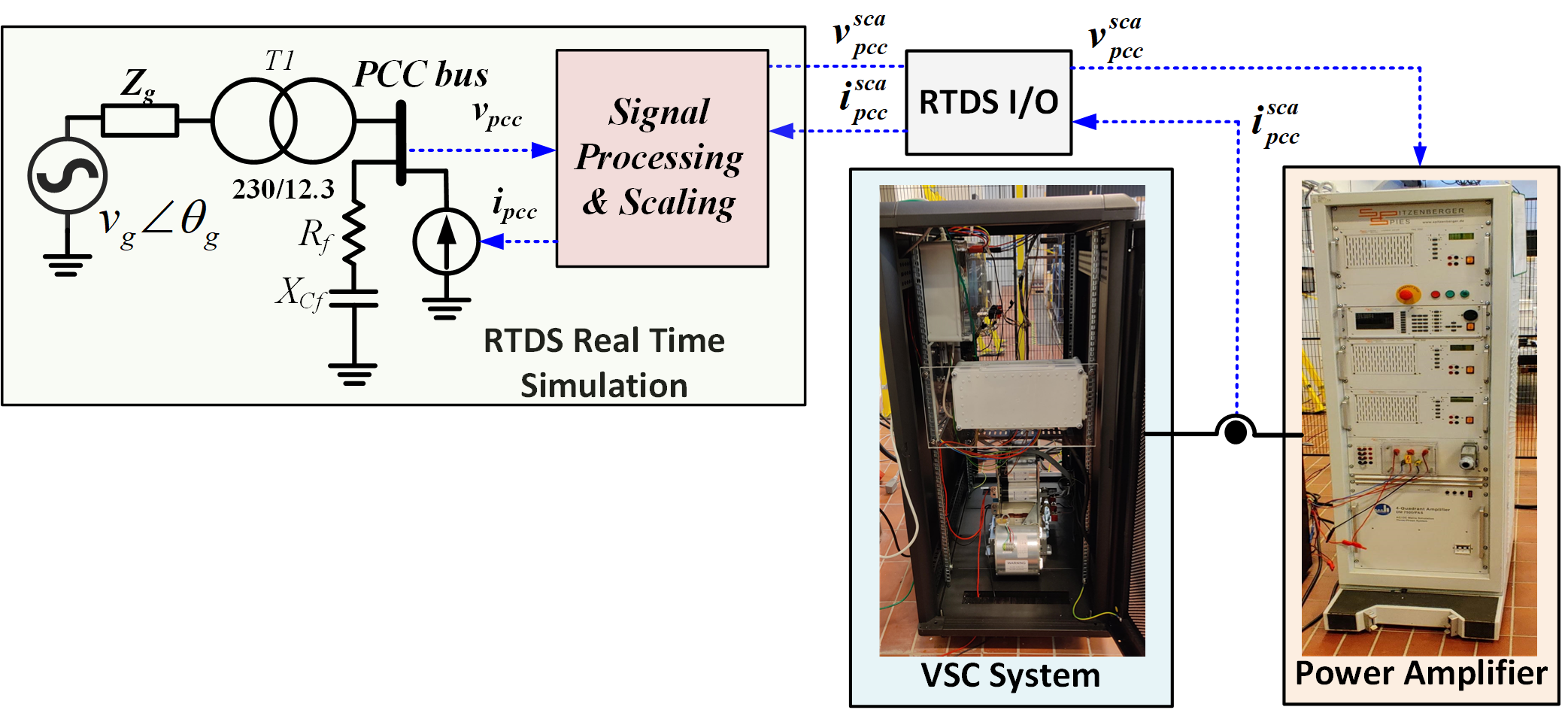}
    \caption{Power hardware in the loop configuration for GFC testing}
    \label{fig:GFC PHIL}
\end{figure}

\begin{figure}[h]
    \centering
    \includegraphics[width=8.5 cm]{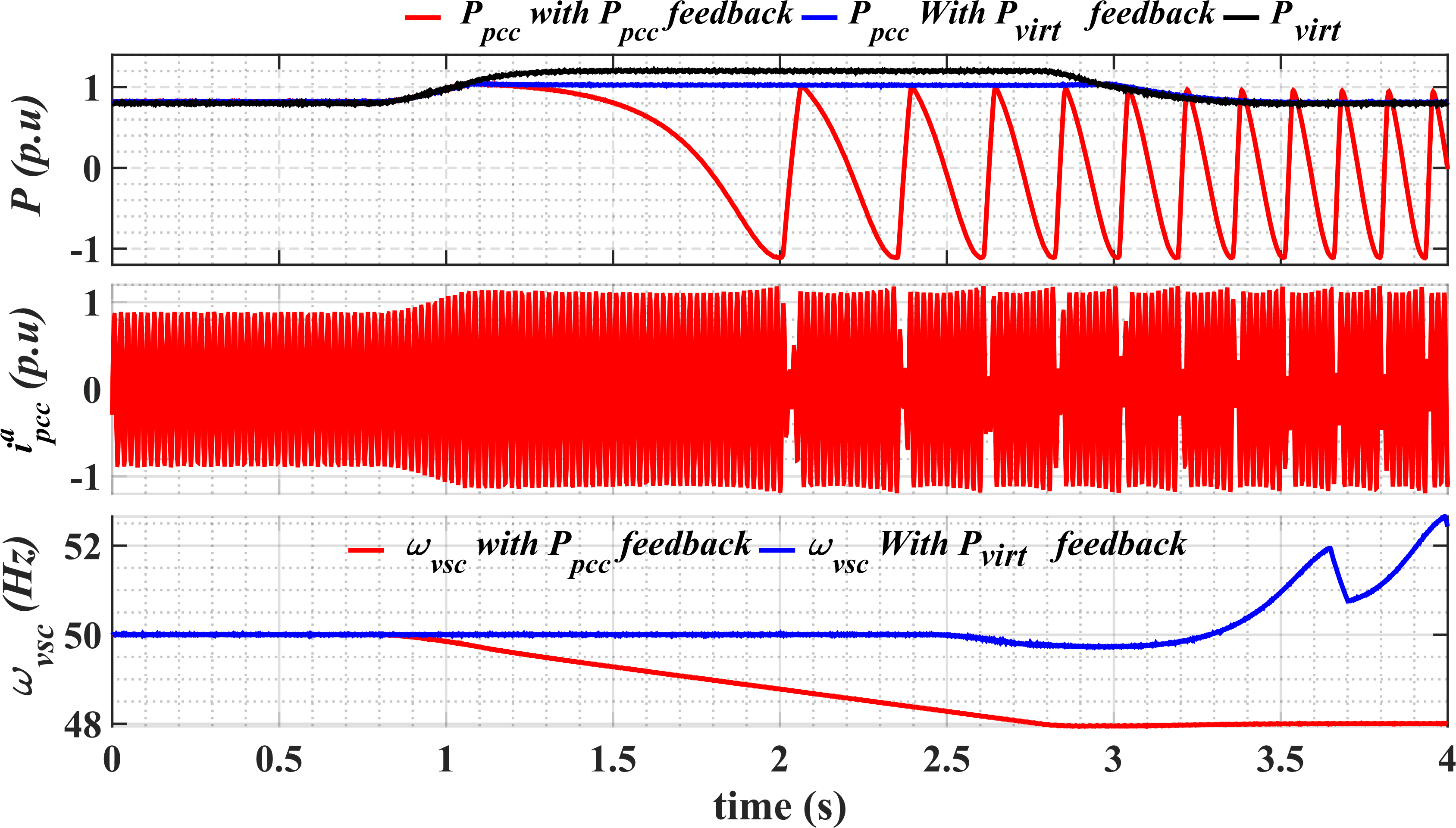}
    \caption{PHIL GFC connected to infinite voltage source, response for a 1 Hz/s ROCOF }
    \label{fig:GFC_rocof_pvirt_PHIL}
\end{figure}

\begin{figure}[h]
    \centering
    \includegraphics[width=8.5 cm]{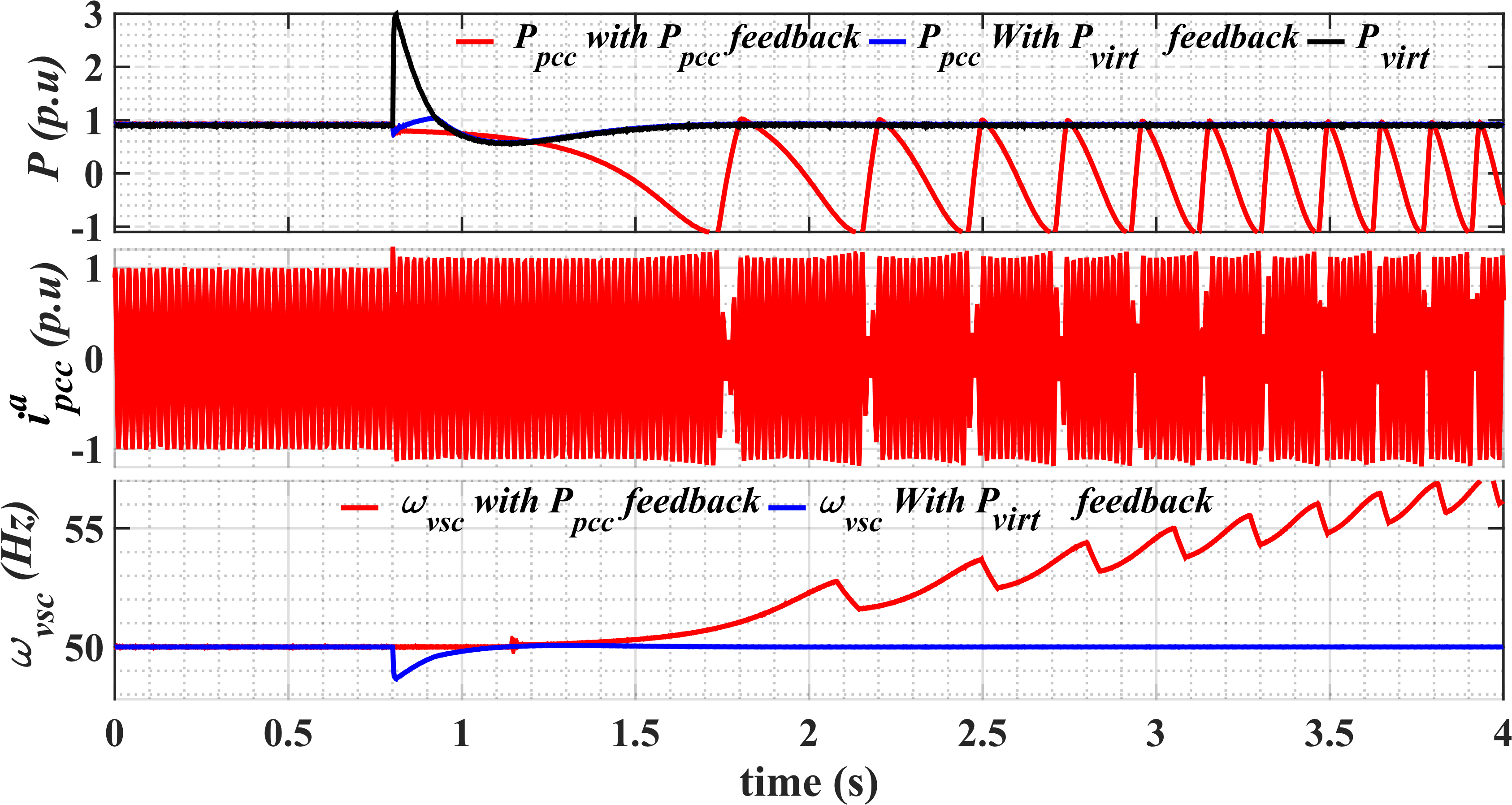}
    \caption{PHIL GFC connected to infinite voltage source, response for a phase jump of 40 deg }
    \label{fig:GFC_phase_shift_40_pvirt_PHIL}
\end{figure}

\begin{figure}[h]
    \centering
    \includegraphics[width=8.5 cm]{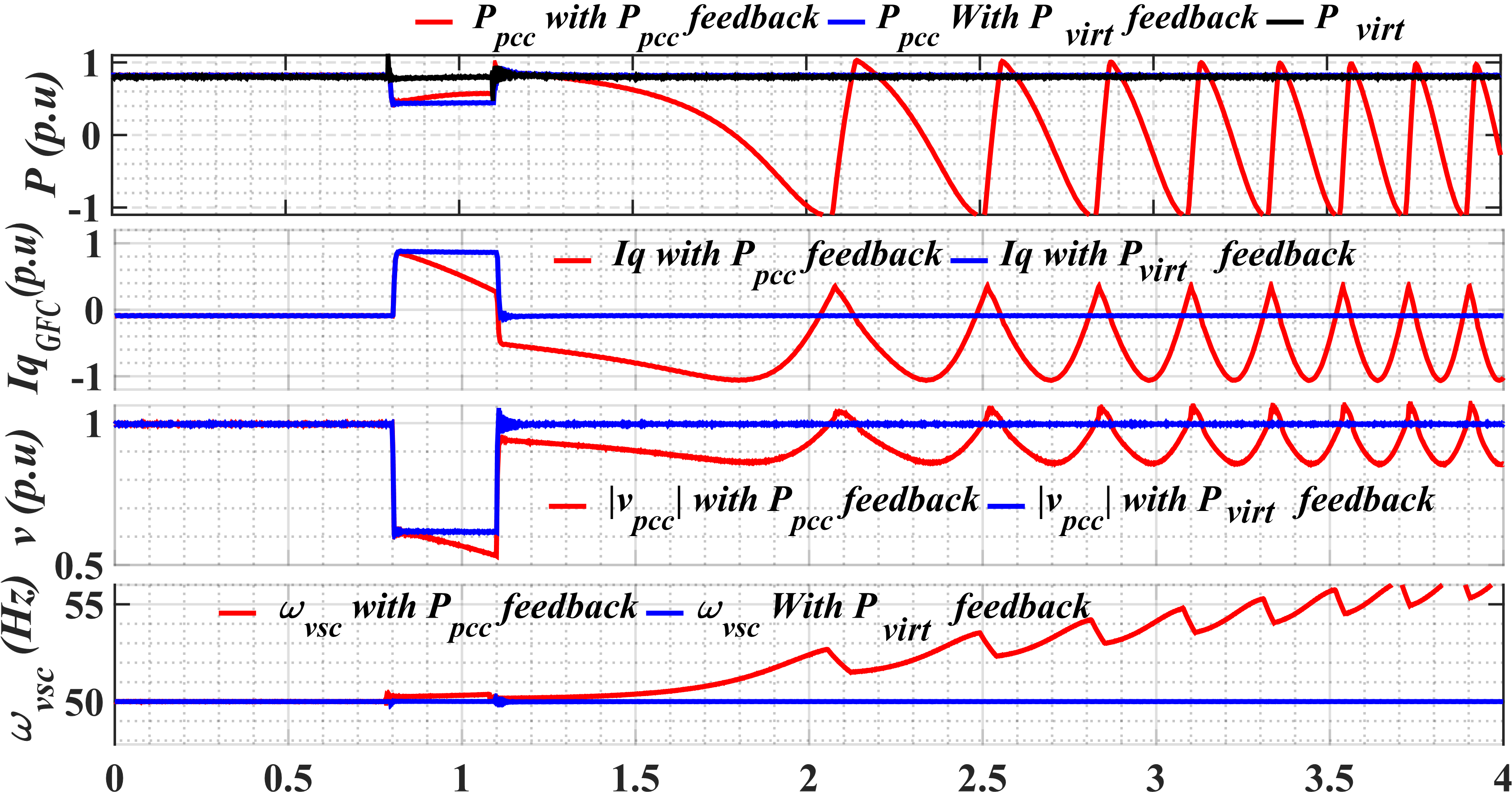}
    \caption{PHIL GFC connected to infinite voltage source, response for a voltage dip of 0.5 pu}
    \label{fig:GFC_fault_pvirt_PHIL}
\end{figure}

\begin{figure}[h]
    \centering
    \includegraphics[width=8.5 cm]{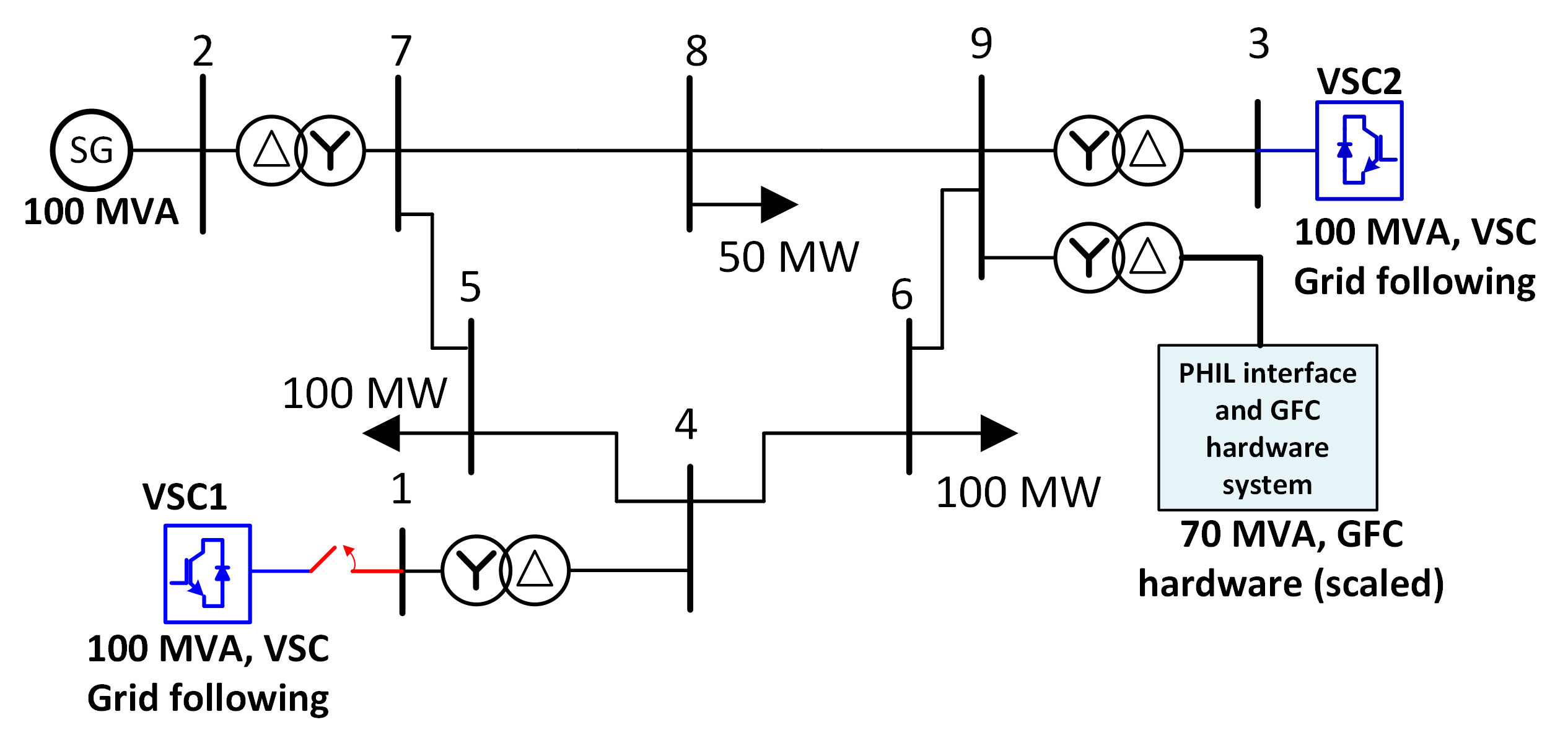}
    \caption{Power hardware in the loop with RTDS simulated modified IEEE-9 bus system }
    \label{fig:PHIL_9_bus}
\end{figure}
\section{Experimental results}
The analysis presented in the previous section is validated through power hardware in the loop (PHIL) simulation. The configuration of the power hardware in the loop study is as shown in Fig. \ref{fig:GFC PHIL}. Firstly, the PHIL study is conducted for GFC connected to an infinite voltage source to validate the analysis presented in section IV. The 230 kV infinite voltage source with Thevenin grid impedance and 230/12.3 kV transformer is simulated in Realtime Digital Simulator (RTDS). The per unit values of the grid impedances and tranformer impedances remain the same as in section IV with a base power of 100 MVA.  A SEMIKRON SkiiP Voltage Source Converter (VSC) stack with an inductive filter and current and voltage sensors and a 2.5 kVA SPITZENBERGER SPIES PAS 2500 linear amplifier are the hardware elements used in the experimental study.  An Ideal Transformer interface method is used to interface the real-time simulation, and the hardware via the amplifier \cite{RTDS}. The RTDS I/O cards are used to exchange the PCC voltage from RTDS to the amplifier and the current coming into the amplifier to the RTDS. This exchange ensures that the VSC is part of the power system network. The voltage and current signal between the RTDS and the hardware is scaled, and the scaling ratio is shown in Table. \ref{table_3_2}. The current feedback signal is conditioned with a first-order low pass filter with a time constant of 250 $\mu s$ to eliminate noise and ensure the stability of the PHIL.

 \begin{figure}[h]
    \centering
    \includegraphics[width=8.5 cm]{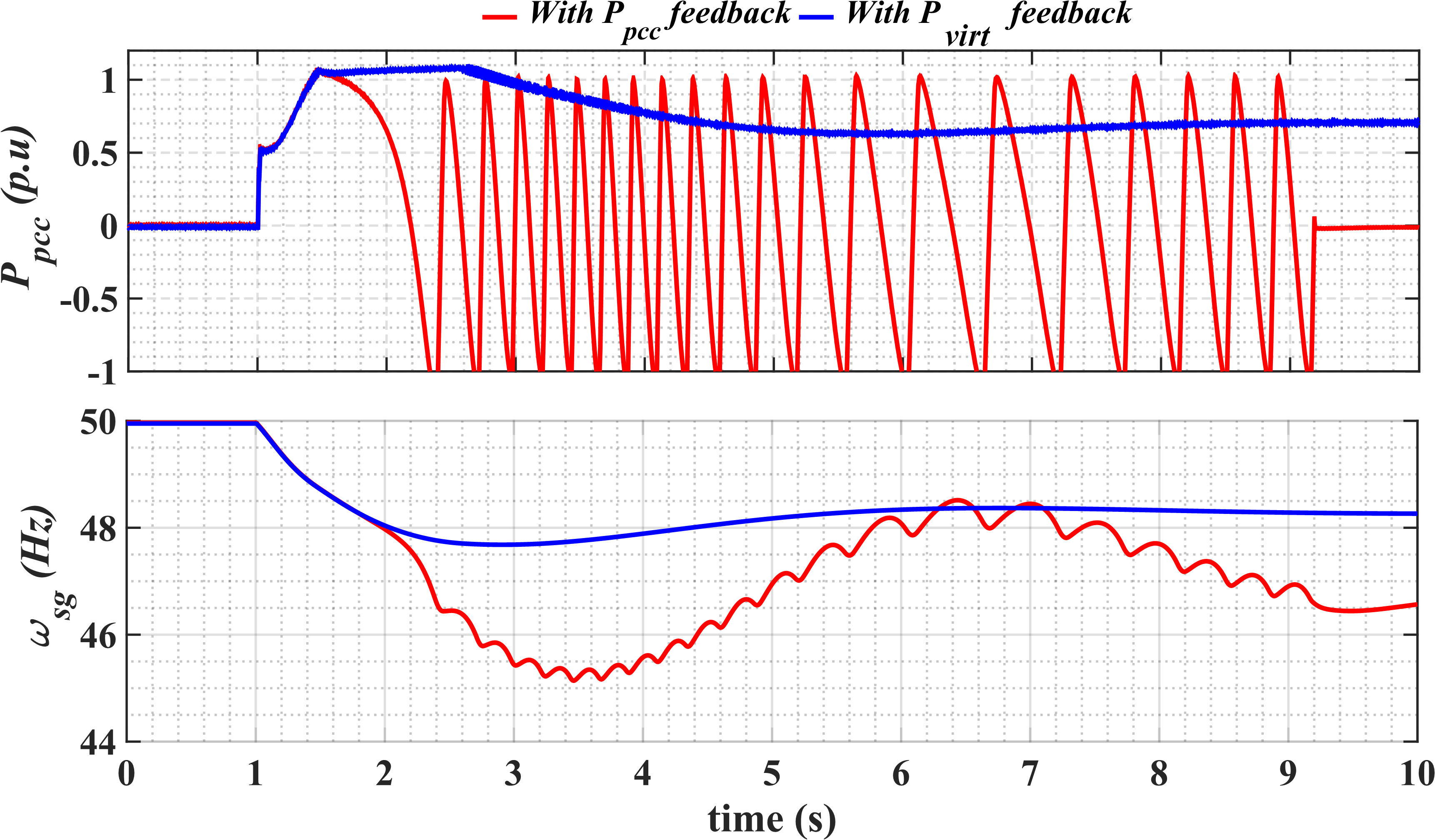}
    \caption{The response from GFC for the VSC1 disconnection event,H is 10 s, $P_{set}$of 0.0 pu, and P-f frequency droop $R_d$ of 5$\%$ }
    \label{fig:GFC_PHIL_9bus1}
\end{figure}
 
 \begin{figure}[h]
    \centering
    \includegraphics[width=8.5 cm]{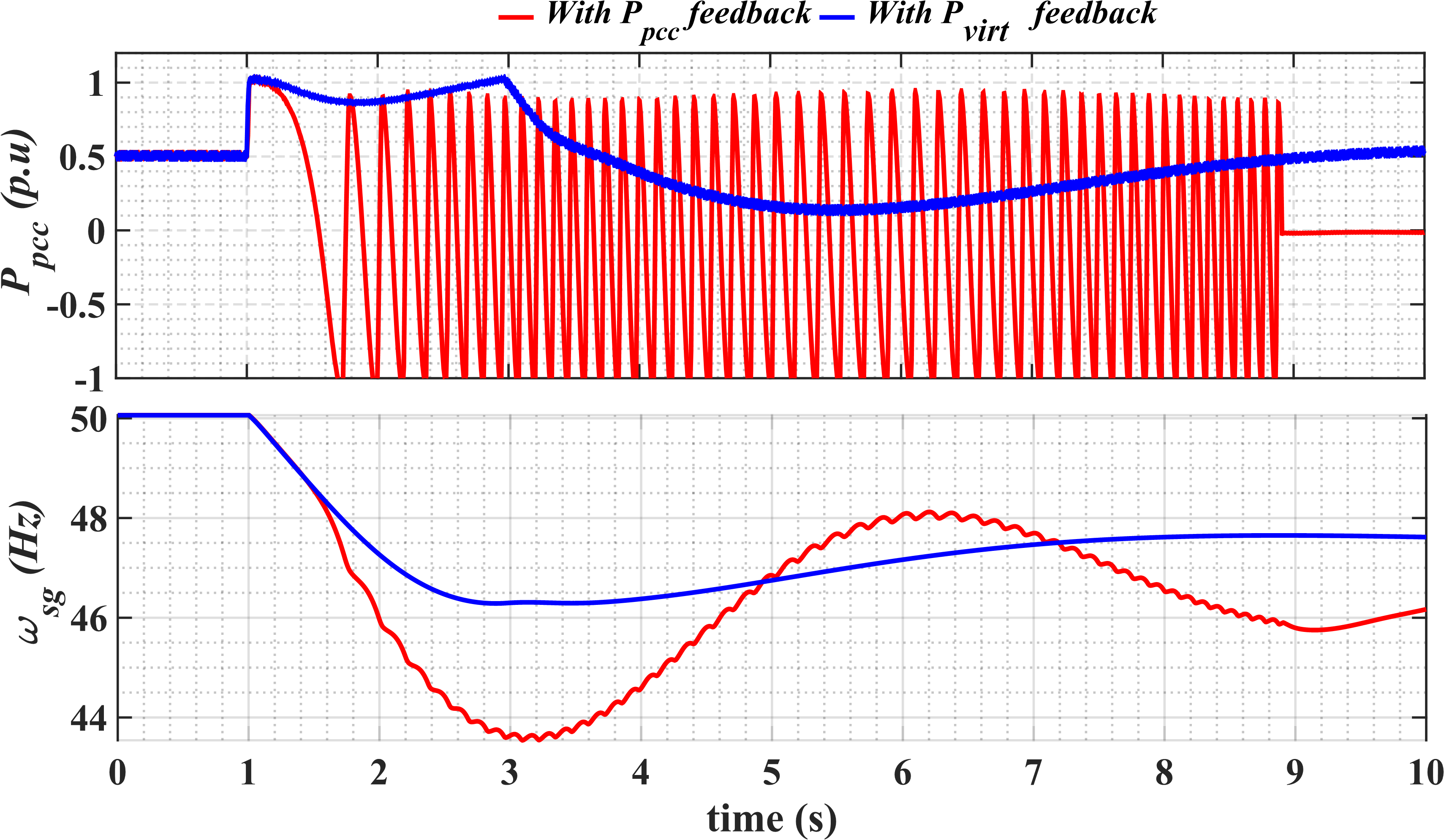}
    \caption{The response from GFC for the VSC1 disconnection event,H is 10 s, $P_{set}$of 0.5 pu, and no P-f frequency droop }
    \label{fig:GFC_PHIL_9bus2}
\end{figure}

The VSC switching frequency is set to 10 kHz, and the inductive filter for the VSC stack is 8 $mH$. The GFC control with a proportional inner current control and the current limit discussed in the paper is implemented on an FPGA-based digital controller from National Instruments (NI). The control is discretized using a trapezoidal integration method with a sampling time of 40 $\mu s$.  The internal control variables such as $P_{virt},\omega_{vsc}$ are captured in the NI controller and transferred to the real-time controller via direct memory access (DMA) first-in-first-out (FIFO) buffers and then saved to a file upon a configurable trigger event. The waveforms at the PCC are captured in the RTDS run time interface. Both sets of data are exported to MATLAB and replotted for enhanced clarity. 
The PHIL results for the infinite bus case are shown in Fig. \ref{fig:GFC_rocof_pvirt_PHIL}-\ref{fig:GFC_fault_pvirt_PHIL}. The PHIL results shown for RoCoF, phase jump, and fault events are in good agreement with the analysis presented in Section IV.  
The reactive current at the PCC ($Iq_{GFC}$) is also plotted for the voltage dip case.  

The PHIL study is expanded to a modified IEEE-9 system representative of future low inertia system with limited SG as shown in Fig. \ref{fig:PHIL_9_bus}. Two of the synchronous generators of the at bus IEEE-9 bus system (at bus 3 and bus 1) are replaced with a commonly used 100 MVA grid following VSC's with active and reactive power control \cite{wen2015analysis}. The hardware GFC is connected to the IEEE 9 bus system at bus 9, the scaling and method of the PHIL interface remain the same as discussed before. The transmission line parameters are the same as the original IEEE 9 bus system \cite{9Bus}. The dispatch power of the SG in the system is adjusted to set the system frequency to 50 Hz. Also, the P-f droop of the SG governer is by default 5$\%$. 
A generation disconnection event of the grid following VSC1 generating 100 MW is considered in this study. Two cases with different droop parameters and dispatch power from grid forming converter are considered in this study. The rotor of the synchronous generator has under damped oscillatory behavior for a load or generation disconnection event. The response from GFC for the VSC1 disconnection event when the programmed inertia constant is 10 s and power setpoint of 0.5 pu is shown in Fig. The active power response of GFC, as well as the rotor speed of the SG($\omega_{sg}$), is shown in Fig. \ref{fig:GFC_PHIL_9bus1}. And the response from GFC for the VSC1 disconnection event when the programmed inertia constant is 10 s and power setpoint of 0.0 pu, and P-f frequency droop $R_d$ of 5$\%$ is shown in Fig.\ref{fig:GFC_PHIL_9bus2} In both cases, GFC, when utilizing measured active power, loses synchronism. In contrast, a seamless entry to the current limiting and a seamless exit from the current limit is achieved when virtual active power is used for synchronization

\section{Conclusion}
 The analysis presented in this paper has revealed the potential transient stability problem for a GFC is greatly accentuated when the GFC enters a current limited operation under differen system events. This paper presents the necessity to evaluate the transient stability problem with transients such as large frequency events, phase jumps, and voltage dips instead of limiting the transient stability analysis just to a fault conditions. A quantitative and illustrative study for the GFC with current control operating against large transient is presented. The paper proposes utilizing the internal virtual power derived from unsaturated current references for ensuring the synchronization under large transients when output GFC current is limited. The results from the Power hardware in the loop (PHIL) experimental tests both on a single GFC connected to an infinite bus as well as on a modified IEEE-9 bus system demonstrates the transient stability challenges for a GFC and the validates of enhanced transient stability of using a virtual power for GFC synchronization.


\vspace{-1.0cm}	
\begin{IEEEbiography}[{\includegraphics[width=1in,height=1.25in,clip,keepaspectratio]{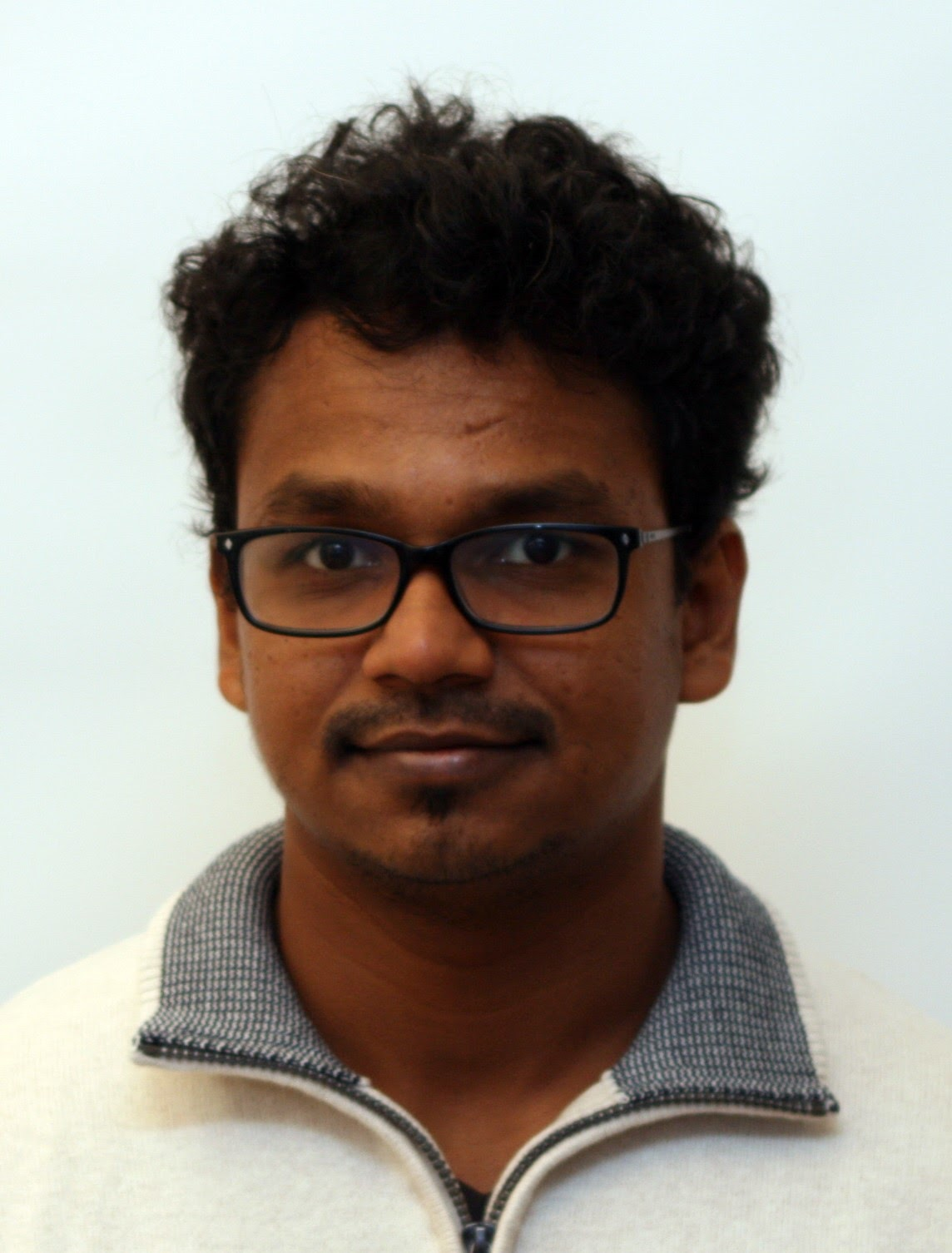}}]
{Kanakesh Vatta KKuni}(S'18) is a PhD student at the Technical University of Denmark (DTU). He received his Master's degree in Electrical Engineering from the Indian Institute of Technology Bombay in the year 2014. From 2016 to 2018, he was Research Engineer in Berkeley Education Alliance for Research in Singapore, while from 2014 to 2016, he worked as a Research Engineer in Energy Management and Microgrid Lab of the National University of Singapore (NUS).His research interests include power system stability, power converter control and microgrids.
\end{IEEEbiography}

\begin{IEEEbiography}[{\includegraphics[width=1in,height=1.25in,clip,keepaspectratio]{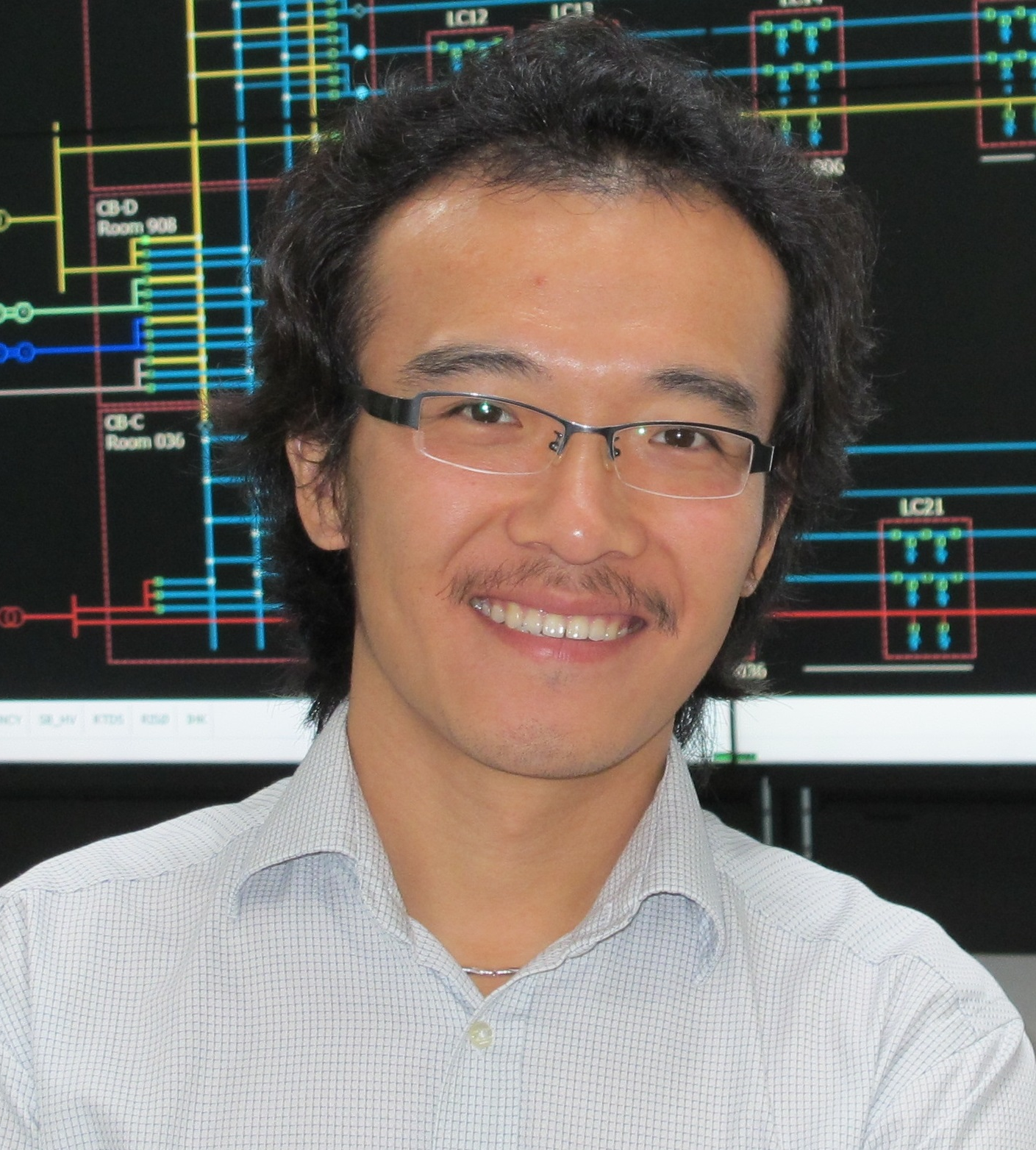}}]
{Guangya Yang} (M'09, SM,14), senior researcher at Technical University of Denmark. He obtained PhD from the University of Queensland, Australia in 2008 afterwards joined Technical University of Denmark as postdoctoral fellow, researcher, and later associate professor. From 2020 to 2021, he was full-time working for Orsted as specialist on electrical design, control, and protection of large offshore wind farms. His research field is security, stability and protection of power systems, with focus on offshore wind applications. He is member of IEC TC88 and TC8, and senior member of IEEE. He has received numerous research grants as principal investigator with the recent H2020 Marie-Curie Innovative Training Networks project InnoCyPES (2021-2025) as coordinator. He is currently leading the editorial board of the IEEE Access Power and Energy Society Section, and is editorial board member of IEEE Transactions on Sustainable Energy, IEEE Transactions on Power Delivery, and Journal of Modern Power System and Clean Energy.
\end{IEEEbiography}

\end{document}